\documentclass[11pt,a4paper]{article}
\usepackage{soul}
\usepackage{jheppub}
\usepackage{natbib}
\usepackage{graphicx}
\usepackage[dvipsnames]{xcolor}

\usepackage{mathtools}
\usepackage{hyperref}
\usepackage{float}
\usepackage{comment}
\usepackage{multirow}
\usepackage{indentfirst}

\def\bea{\begin{eqnarray} }
\def\eea{ \end{eqnarray} } 

\newcommand{\circled}[1]{\raisebox{.5pt}{\textcircled{\raisebox{-.9pt} {#1}}}}

\definecolor{Blu}{rgb}{0.,0.,1.}

\newcommand{\Mp}{M_{PL}}
\newcommand{\crit}{\rm{crit}}

\newcommand{\rhoBBNT}{\rho_{\phi}(T_{\rm{BBN}})}

\def\GW{\text{GW}}
\def\MpR{M_{PL}}

\newcommand{\eq}[1]{Eq.~\eqref{#1}}

\author[a]{Anirban Biswas,}
\author[b]{Arpan Kar,}
\author[b,c]{Bum-Hoon Lee,}
\author[b,c]{Hocheol Lee,}
\author[b]{Wonwoo Lee,}
\author[b,c]{Stefano Scopel,}
\author[b,c]{Liliana Velasco-Sevilla,}
\author[d,e]{Lu Yin}
\affiliation[a]{Department of Physics \& Lab of Dark Universe, Yonsei University, Seoul 03722, Republic of Korea}
\affiliation[b]{Center for Quantum Spacetime, Sogang University, Seoul 121-742, South Korea}
\affiliation[c]{Department of Physics, Sogang University, Seoul 121-742, South Korea}
\affiliation[d]{Department of Physics, Shanghai University, Shanghai 200444, China}
\affiliation[e]{Asia Pacific Center for Theoretical Physics (APCTP)
San 31, Hyoja-dong, Nam-gu, Pohang 790-784, South Korea}
\emailAdd{anirban.biswas.sinp@gmail.com}
\emailAdd{arpankarphys@gmail.com}
\emailAdd{bhl@sogang.ac.kr}
\emailAdd{insaying@sogang.ac.kr}
\emailAdd{warrior@sogang.ac.kr}
\emailAdd{scopel@sogang.ac.kr}
\emailAdd{liliana.velascosevilla@gmail.com}
\emailAdd{lu.yin@apctp.org}

\title{Gauss-Bonnet Cosmology: large--temperature behaviour and bounds from Gravitational Waves}

\abstract{We provide a transparent discussion of the high temperature asymptotic behaviour of Cosmology in a dilaton-Einstein-Gauss-Bonnet (dEGB) scenario of modified gravity with vanishing scalar potential. In particular, we show that it has a clear interpretation in terms of only three attractors (stable critical points) of a set of autonomous differential equations: $w=-\frac{1}{3}$, $w=1$ and $1<w<\frac{7}{3}$, where $w\equiv p/\rho$ is the equation of state, defined as the ratio of the total pressure and the total energy density. All the possible different high--temperature evolution histories of the model are exhausted by only eight paths in the flow of the set of the autonomous differential equations. Our discussion clearly explains why five out of them are characterized by a swift transition of the system toward the attractor, while the remaining three show a more convoluted evolution, where the system follows a meta--stable equation of state at intermediate temperatures  before eventually jumping to the real attractor at higher temperatures. Compared to standard Cosmology, the regions of the dEGB parameter space with $w=-\frac{1}{3}$ show a strong enhancement of the expected Gravitational Wave stochastic background produced by the primordial plasma of relativistic particles of the Standard Model. This is due to the very peculiar fact that dEGB allows to have an epoch when the energy density $\rho_{\rm rad}$ of the relativistic plasma dominates the energy of the Universe while at the same time the rate of dilution with $T$ of the total energy density is slower than what usually expected during radiation dominance. This allows to use the bound from Big Bang Nucleosynthesis (BBN) to put in dEGB a constraint $T_{\rm RH}\lesssim (10^8 - 10^9)$ GeV on the reheating temperature of the Universe $T_{\rm RH}$. Such BBN bound is complementary to late-time constraints from compact binary mergers.
}

\begin{document}
\preprint{CQUeST-2024-0735}
\maketitle

\section{Introduction}
\label{sec:introduction}

In spite of its success in describing the observed Universe, General Relativity (GR) is believed to be incomplete, because it is difficult to reconcile with the other fundamental interactions of the Standard Model of Particle Physics (SM). Cosmology and astrophysics provide useful approaches to explore its possible extensions. In a previous paper~\cite{GB_WIMPS_sogang}, we analyzed a specific example of such extensions: the dilaton-Einstein-Gauss-Bonnet (dEGB) scenario, obtained by adding to the Einstein action a specific quadratic combination of the curvature non-minimally coupled to a scalar field~\cite{Hwang:1999gf, Satoh:2008ck}. 

In extensions of Einstein Gravity such as string theory higher curvature terms are expected to appear and to become important in the early Universe. dEGB represents the simplest example with a higher curvature term of Horndeski’s theory, the most general scalar-tensor theory having equations of motion with second-order time derivatives in four-dimensional spacetime~\cite{Horndeski:1974wa}, in which the theory does not have a ghost state~\cite{Woodard:2015zca}. As a consequence, dEGB 
provides a very effective and popular framework to address both theoretical and phenomenological issues beyond General Relativity~\cite{Boulware:1985wk, Zwiebach:1985uq, Cai:2001dz, Nojiri:2006ri, Nojiri:2010wj, Clifton:2011jh, Harko:2013gha, Bahamonde:2017ize, Odintsov:2018ggm, Alexander:2019rsc, Banerjee:2020xcn}.


In Cosmology dEGB has been extensively studied during the inflationary era~\cite{Hwang:1999gf, Satoh:2008ck, Guo:2010jr, Koh:2014bka, Koh:2014fxg, Lahiri:2016jqv, Fomin:2017vae, Yi:2018gse, Odintsov:2018zhw, Pozdeeva:2020apf, Kawai:2021edk, Nojiri:2023jtf, Kawai:2023nqs}, in the reheating period at the end of inflation ~\cite{vandeBruck:2016xvt, Koh:2018qcy, Rashidi:2020wwg, Oikonomou:2024jqv}, and has been used to explain the accelerated expansion observed in the late--time Universe~\cite{Nojiri:2005vv, Cognola:2006sp, Nojiri:2023mvi, MohseniSadjadi:2023cjd, TerenteDiaz:2023iqk}. In Astrophysics dEGB has been used to study black holes with a scalar hair~\cite{Kanti:1995vq, Guo:2008hf, Ahn:2014fwa, Khimphun:2016gsn, Antoniou:2017acq, Doneva:2017bvd, Silva:2017uqg, Myung:2018iyq, Lee:2018zym, Lee:2021uis, Papageorgiou:2022umj, Hyun:2024sfv}, and other objects~\cite{Lee:2016yaj, Chew:2020lkj, Lee:2021nwg}. The dEGB theory has been also tested against observed gravitational waves (GW) signals from black hole-black hole (BH-BH) or black hole-neutron star (BH-NS) merger events~\cite{Nair:2019iur, Okounkova:2020rqw, Wang:2021jfc, Perkins:2021mhb, BH-NS_GB_2022}.

Specifically, in Ref.~\cite{GB_WIMPS_sogang} we studied in detail the thermal evolution of a Cosmological model where GR is modified by the dEGB term and the potential of the scalar field vanishes. In particular, we focused on a range of temperatures between BBN and a few TeV, with the goal to constrain the specific scenario of the thermal decoupling of a Weakly Interacting Massive Particle (WIMP). 
Indeed, while we have no direct probe of its Universe expansion rate, the epoch between reheating and Big Bang Nucleosynthesis
is crucial to shed light on physics beyond the SM such as the nature of Dark Matter~\cite{Salati:2002md, Rosati:2003yw, Kang:2008zi, Capozziello:2012uv, Capozziello:2015ama, Meehan:2015cna, Lambiase:2016log, profumo_relentless_2017} or the origin of the baryon asymmetry in the Universe~\cite{Buchmuller:2005eh,Kawai:2017kqt}.

Interestingly, the numerical analysis of Ref.~\cite{GB_WIMPS_sogang} unveiled that, in spite of a high degree of non linearity and phenomenological complexity at low temperatures, at large-enough temperatures dEGB exhibits instead only very few asymptotic behaviours, characterized by the equation of state $w=p/\rho$ (where $\rho$ and $p$ are the total energy density and  the pressure of the Universe, respectively), when the model is required to reproduce Standard Cosmology at BBN. Specifically, the study of Ref.~\cite{GB_WIMPS_sogang} showed that within dEGB the asymptotic value at high temperature of the equation of state falls only into three cases: $w=1$, $w=-1/3$, $1 \le w \lesssim 2.3$.  Clearly, such behaviour suggests a pattern of attractor solutions emerging at high temperatures. However, the numerical approach that we used in Ref.~\cite{GB_WIMPS_sogang} to solve the coupled Friedmann differential equations of the system did not provide a transparent interpretation of such pattern, that also required to extend the analysis beyond the temperature relevant for the WIMP decoupling effect that we wished to discuss. As a consequence, in Ref.~\cite{GB_WIMPS_sogang} we did not attempt to provide a systematic discussion of the asymptotic behaviour of dEGB at high temperatures. The goal of the present paper is to fill this gap, extending the analysis of dEGB to higher temperatures and providing a transparent and systematic discussion of its high temperature asymptotic behaviour. The main result of the present analysis is twofold:

\begin{itemize}
    \item An appropriate change of variables~\cite{Koh:2014bka} allows to express the evolution of the dEGB Friedmann equations in terms of a set of two autonomous coupled differential equations whose critical points and attractors can be studied in a  straightforward way, providing a clear and transparent understanding of the pattern of the high--temperature asymptotic behaviours that emerged from the numerical analysis of Ref.~\cite{GB_WIMPS_sogang}. In particular, such interpretation allows also to understand that one of the regimes identified as asymptotic in Ref.~\cite{GB_WIMPS_sogang} was indeed meta--stable, i.e. would eventually converge to one of the other possible asymptotic regimes at higher temperatures (such feature was irrelevant for the WIMP analysis of Ref.~\cite{GB_WIMPS_sogang}).

    \item In the regions of the dEGB parameter space where the asymptotic behaviour of the equation of state is $w=-1/3$ the potential signal of stochastic GW produced by the thermal plasma of the Standard Model particles in the early Universe~\cite{Ghiglieri:2015nfa,Ghiglieri:2020mhm} can exceed the Big Bang Nucleosynthesis (BBN) bounds for values of the reheating temperature as low as $T_{RH}\simeq 10^9$ GeV. This is at variance with Standard Cosmology, where for $T_{RH}$ as high as $10^{16}$ GeV such contribution is diluted away and negligible at the time of BBN, and provides some prospects to detect such signal experimentally in future detectors sensitive to the high--frequency GW spectrum~\cite{Ito:2019wcb, Herman:2022fau}.  
\end{itemize}

Our paper is organized as follows. In Section~\ref{sec:GB_theory} we outline the dEGB scenario of modified gravity and fix our notations; in Section~\ref{sec:numerical} we extend to $T\lesssim10^{16}$ GeV the discussion of the numerical solutions of the Friedmann equations in
dEGB Cosmology of Ref.~\cite{GB_WIMPS_sogang}; in Section~\ref{sec:tracking} we re-express the Friedmann equations discussed in Section~\ref{sec:numerical} in terms of a system of two coupled autonomous differential equations and provide a detailed discussion of its critical points and attractors that allows to interpret the numerical results of Section~\ref{sec:numerical} in a transparent way (in particular such discussion is summarized schematically in Figure~\ref{fig:critical_points} and in Table~\ref{table:critical_points}). Section~\ref{sec:GW_constraints} is devoted to the GW bounds, with Section~\ref{sec:GW_stochastic} dealing with the stochastic background and Section~\ref{sec:GW_binaries} with Late Universe constraints from BH-BH and BH-NS merger events. In particular, in Section~\ref{sec:GW_theory} we summarize how the GW stochastic signal is calculated, and in Section~\ref{sec:GW_experiment} we discuss its possible observable implications. Finally, we combine the observational constraints on dEGB Cosmology from GW observables in Section~\ref{sec:results}, and provide our Conclusions in Section~\ref{sec:conclusions}.
For completeness, in Appendix~\ref{app:thermalcorr} we summarize the thermal corrections calculated in Ref.~\cite{Ghiglieri:2020mhm} that are used in Section~\ref{sec:GW_theory} to calculate the source term of the stochastic GW background.

\section{Summary of the dilaton-
Einstein-Gauss-Bonnet scenario}
\label{sec:GB_theory}

In order to fix our notation, in this Section we briefly outline the dEGB scenario that we analyzed in~\cite{GB_WIMPS_sogang}. More details can also be found in~\cite{Kanti:1995vq, Koh:2014bka,  GB_WIMPS_sogang}. The action is given by

\begin{equation}
S=  \int_{\mathcal M} \sqrt{-g} \ d^4 x \left[ \frac{R}{2\kappa}
-\frac{1}{2}{\nabla_\mu}\phi {\nabla^\mu}\phi -V(\phi) + \ f(\phi)R^2_{\rm GB} + {\mathcal L}^{\rm rad}_{m} \right]\,,
\label{action}
\end{equation} 
\noindent where $\kappa \equiv 8\pi G =1/M^2_{\rm PL}$ (where $M_{\rm PL}$ is the reduced Planck mass), $R$ is the Ricci scalar of the spacetime ${\cal M}$, 
$R^{2}_{\rm GB} = R^2 - 4 R_{\mu\nu}R^{\mu\nu} + R_{\mu\nu\rho\sigma} R^{\mu\nu\rho\sigma}$ is the Gauss-Bonnet term and ${\mathcal L}^{\rm rad}_{m}$ describes radiation (i.e. all the relativistic species). Following the same approach that we adopted in Ref.~\cite{GB_WIMPS_sogang}  we will assume $V=0$. The function $f(\phi)$ is a coupling function between the scalar field and the Gauss-Bonnet term for which we assume 
\bea
f(\phi) = \alpha e^{\gamma\phi},
\label{eq:f_GB}
\eea
\noindent where $\alpha$ and $\gamma$ are constants. An exponential form for $f$ arises naturally within theories where gravity is coupled to the dilaton~\cite{Kanti:1995vq}. In particular, although in string theory the natural sign of the $\alpha$ coefficient is positive~\cite{Boulware:1985wk}, in Refs.~\cite{Koh:2014bka, Lee:2018zym, Lee:2021uis} it was shown that for both signs of $\alpha$ black-hole solutions can be found. In our phenomenological analysis we will adopt both signs.

By varying the action the equation of motion for the scalar field  $\square \phi + f' R^{2}_{\rm GB} =0\,,$
and Einstein's equations
$R_{\mu\nu}-\frac{1}{2}g_{\mu\nu}R =\kappa T^{\rm tot}_{\mu\nu}$ can be obtained, where $T^{\rm tot}_{\mu\nu} = T^{\phi}_{\mu\nu} + T^{\rm GB}_{\mu\nu} + T^{\rm rad}_{\mu\nu}$ is given by three contributions, corresponding to the scalar field, the GB term and radiation. In particular,  $T^{\rm tot}_{\mu\nu}$ satisfies the continuity equation, although due to the interaction between the Gauss-Bonnet term and the scalar field the two components $T^{\phi}_{\mu\nu}$ and $T^{\rm GB}_{\mu\nu}$ do not satisfy it separately.

The evolution of the universe is well described by homogeneous and isotropic space on a large scale. We take the spatially flat Friedmann-Lema\^{i}tre-Robertson-Walker (FLRW) metric, i.e.,
\begin{equation}
ds^2 = - dt^2 + a^2(t)\delta_{ij} dx^i dx^j\,, \label{frwmewtric}
\end{equation}
\noindent in terms of the scale factor $a(t)$. As a consequence, the scalar field depends only on time, $\phi=\phi(t)$. From the energy-momentum tensor one can then obtain the  contributions $\rho_i$ and $p_i$ to the energy density and the pressure, with $i$ = $\phi$, $\rm{GB}$, $\rm{rad}$. They enter the Friedmann equations in the familiar form: 
\begin{eqnarray}
H^2 &=& \frac{\kappa}{3} \rho\,,
\label{Eqfrw1}
\\
{\dot H} &=& - \frac{\kappa}{2} (\rho+p) \,,
\label{Eqfrw2} \\
{\ddot \phi} &+& 3H {\dot \phi} + V_{\rm GB}'=0\,,
\label{Eqfrw3}
\end{eqnarray}

\noindent where $\rho = \rho_{\phi} + \rho_{\rm GB} +\rho_{\rm rad}$, $p = p_{\phi} + p_{\rm GB} +p_{\rm rad}$,
and, each component has the following explicit form:
\begin{eqnarray}
\rho_{\phi} &= & p_{\phi}=\frac{1}{2}{\dot{\phi}}^2,\\
\rho_{\rm GB} &=& -24\dot{f} H^3 = -24f^\prime{\dot{\phi}} H^3=-24{\alpha}\gamma e^{\gamma\phi}\dot{\phi}H^3,
\label{eq:rho_GB}
\\
p_{\rm GB} &=& 8 \left(f^{\prime\prime}\dot{\phi}^2+f^\prime\ddot{\phi}\right) H^2 + 16 {f^\prime}\dot{\phi} H(\dot{H}+H^2)\nonumber\\
&=& 8\frac{d(\dot{f}H^2)}{dt}+16 \dot{f}H^3 =8\frac{d(\dot{f}H^2)}{dt} -\frac{2}{3} \rho_{\rm GB}\,,
\label{eq:p_GB}
\\
\rho_{\rm rad} &=& 3 p_{\rm rad}=\frac{\pi^2}{30}\,g_{*}\,T^4,
\label{eq:p_rad}
\end{eqnarray}
where $g_{*}$ is the number of effective 
relativistic degrees of freedom in equilibrium with the thermal bath. Notice that, as far as the evolution of the Hubble constant $H$ is concerned, in Eqs.~(\ref{Eqfrw1}) and (\ref{Eqfrw2}) $\rho$ and $p$ can be interpreted as the total energy density and the pressure of the Universe with $\rho$ positive--defined. On the other hand, ${\rho}_{\rm GB}$ and $p_{\rm GB}$ in Eqs.~(\ref{eq:rho_GB}) and (\ref{eq:p_GB}), which can be interpreted as the effective energy density and pressure coming from the GB term, are not necessarily positive. Moreover, in Eq.~(\ref{Eqfrw3}) $V^{\prime}_{\rm GB}$ is 
\begin{eqnarray}
V^{\prime}_{\rm GB}&\equiv& -f' R^{2}_{\rm GB}= -24 f' H^2({\dot H} + H^2)=24 {\alpha}\gamma e^{\gamma\phi} q H^4,
\label{eq:v_gb_prime}
\end{eqnarray}
which is technically not the gradient of a potential, but just an excess of notation to indicate that it drives the scalar field evolution. In~Eq.(\ref{eq:v_gb_prime}) 
$q=-\frac{\ddot{a}a}{\dot{a}^2} =\frac{1}{2} (1+3w)$ is the usual deceleration parameter.



\section{Numerical solutions of the Friedmann equations}
\label{sec:numerical}

In this Section, we solve Eqs.(\ref{Eqfrw2}) and (\ref{Eqfrw3}) numerically, after writing them as a set of three first order coupled differential equations, with the goal to investigate their high--temperature behaviour. Throughout the paper we present our numerical results in geometric units defined by $\kappa = 8 \pi G = 1$, $c = 1$\footnote{In relativistic units of $c = 1$, the dimension of $\kappa = 8 \pi G$ is $[\rm length] \over [\rm mass]$. Also, $[\alpha]=[\rm mass][\rm length]$, $[\gamma]=1/[\phi]=[\sqrt{\kappa}]$. All the quantities can be written in terms of length units by appropriate factors of $\kappa$. In geometric units $\alpha$ is in km$^2$ while $\phi$ as well as $\gamma$ are dimensionless.}.

Assuming iso-entropic expansion ($sa^3$ = constant with $s = (2 \pi^2 / 45) g_{*s} T^3$ the entropy density and $g_{*s}(T)$, the number of entropy degrees of freedom), the change of variable from time to temperature is given as usual as
\begin{eqnarray}
\dfrac{dT}{dt} = -\dfrac{H\,T}{\beta}\,,
\label{eq:T-t}
\end{eqnarray}
\noindent where
\begin{eqnarray}
\beta = \left(1+\dfrac{1}{3}\dfrac{d \ln g_{*s}}{d \ln T}\right)\,,
\label{eq:beta_T_t}
\end{eqnarray}
and before neutrino decoupling $g_{*s}$ = $g_{*}$~\cite{g_T_Steigman2012}. We fix the boundary conditions at the temperature of Big Bang Nucleosynthesis, $T_{\rm BBN} = 1$ MeV, by setting $\phi(T_{\rm BBN})$  $\equiv \phi_{\rm BBN}$,  $\dot{\phi}(T_{\rm BBN})$  $\equiv \dot{\phi}_{\rm BBN}$ and $H(T_{\rm BBN})$ $\equiv H_{\rm BBN}$ and evolve the differential equations ``backward in time" to higher temperatures. Since the solutions of the Friedmann equations become invariant under a simultaneous change of sign of ${\phi}_{\rm BBN}$, $\dot{\phi}_{\rm BBN}$ and $\gamma$, without loss of generality in our analysis we fix $\dot{\phi}(T_{\rm BBN})>0$ and study both signs of $\gamma$. We stop our integration at $T\simeq 10^{16}$ GeV, since we assume that the thermal plasma is kept in thermal equilibrium by standard perturbative interactions, which are frozen out and are ineffective at higher temperatures~\cite{Kolb_Turner_1990}.

As far as the boundary condition for $\phi$ is concerned, we note that a shift of $\phi_{\rm BBN}$ is equivalent to a redefinition of the $\alpha$ parameter
\begin{equation}
{\phi^{\prime}}_{\rm BBN} = \phi_{\rm BBN} + \phi_{0},\,\,\,\alpha^{\prime} = \alpha \hspace{0.5mm} e^{-\gamma \phi_{0}},\,\,\,\gamma^{\prime} = \gamma.
\label{eq:Gauge}
\end{equation}
\noindent In other words, any choice of $\phi_{\rm BBN}$ corresponds to a
specific gauge fixing and the quantity
\begin{equation}
\tilde{\alpha}=\alpha \hspace{0.5mm} e^{\gamma \phi_{\rm BBN}},
\label{eq:alpha_tilde}
\end{equation}
\noindent is invariant under the gauge transformation~(\ref{eq:Gauge}).
In our analysis, we will show our results in terms of $\tilde{\alpha}$ (which is equivalent to adopt the gauge $\phi_{\rm BBN}= 0$)\footnote{Indeed, $\phi$ is not a degree of freedom of the equivalent system of autonomous equations discussed in Section~\ref{sec:autonomous}, and only appears when fixing the boundary conditions at BBN.}.


The ratio of $\rhoBBNT=\frac{1}{2}\dot{\phi}_{\rm BBN}^2$ to the total energy density at BBN, $\eta=\frac{\rhoBBNT}{\rho(T_{\rm BBN})}$, is constrained by the upper bound on the effective number of neutrino flavors $N_{eff}\le$ 2.99 $\pm$ 0.17~\cite{planck_2018}. This bound is then translated to a bound on $\eta$: $\eta \le 3 \times 10^{-2} $. We will adopt three benchmarks for $\dot{\phi}_{\rm BBN}$ corresponding to $\eta =0 $, the upper bound $\eta =3\times 10^{-2}$ and an illustrative intermediate value, $\eta = 10^{-4}$. 

The boundary condition on $H(T_{\rm BBN})$ = $H_{\rm BBN}$ can be obtained by solving  Eq. (\ref{Eqfrw1}) at $T_{\rm BBN}$ and taking the solution closer to $H_{\rm rad}(T_{\rm BBN})$. At fixed $\eta$, the $\dot{\phi}_{\rm BBN}$ linear factor that appears in the expression of $\rho_{\rm GB}$  can be expressed in terms of $H_{\rm BBN}$ through $\dot{\phi}_{\rm BBN}$ = $\sqrt{2\rho_{\phi}}$ = $\sqrt{6\eta/\kappa}H_{\rm BBN}$,
where the sign of the square root is positive because we take $\dot{\phi}_{\rm{BBN}} > 0$. Then Eq. (\ref{Eqfrw1}) is equivalent to the following quadratic equation in $H^2_{\rm BBN}$:
\begin{equation}
8 \sqrt{6\kappa\eta} f^{\prime}(0) H_{\rm BBN}^4+(1-\eta) H_{\rm BBN}^2-\frac{\kappa}{3} \rho_{\rm rad}(T_{\rm BBN})
=0\label{eq:quadratic_H}. \\ 
\end{equation}

\begin{figure*}[htb!]
\centering
\includegraphics[width=7.49cm,height=6.1cm]{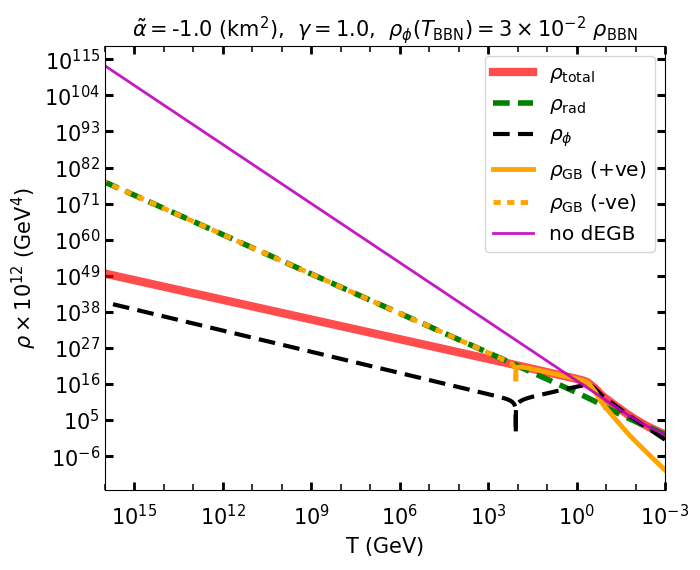}
\includegraphics[width=7.49cm,height=6.1cm]{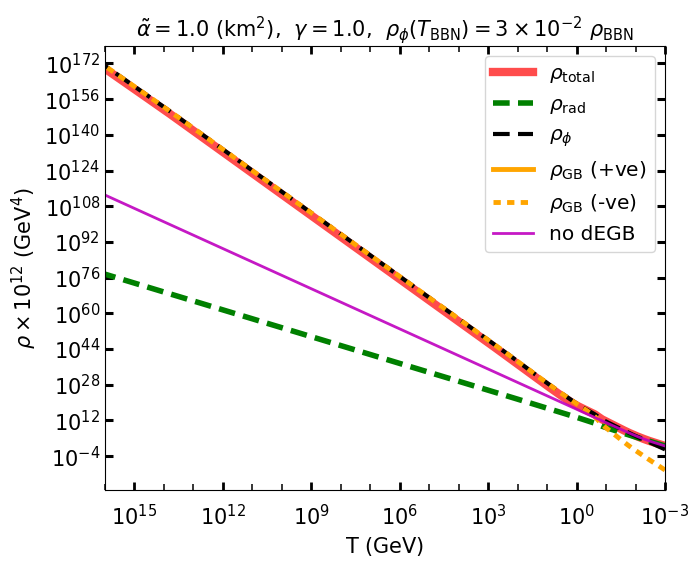}
\includegraphics[width=7.49cm,height=6.1cm]{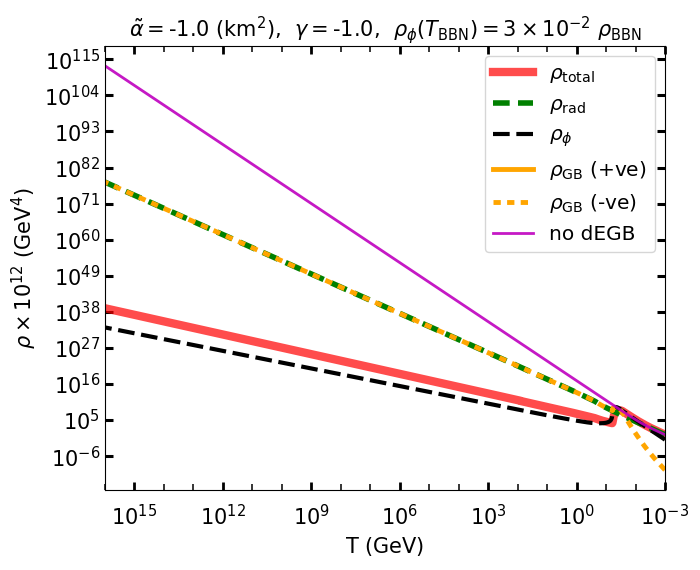}
\includegraphics[width=7.49cm,height=6.1cm]{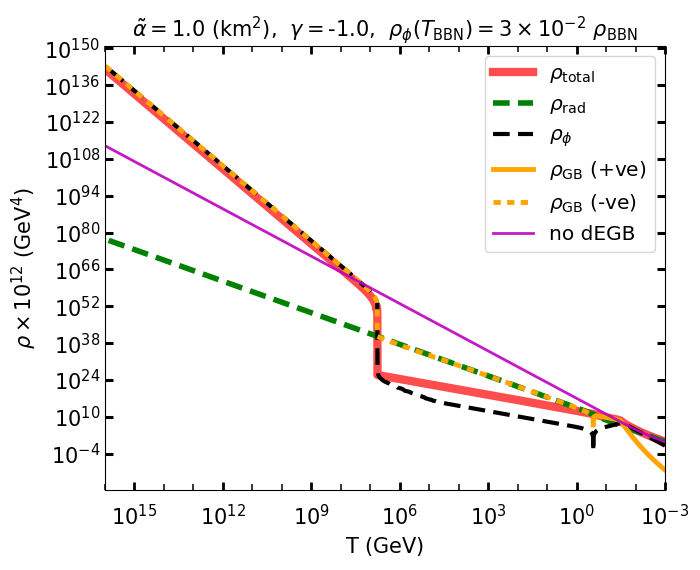}
\caption{Evolution of $\rho_{\rm rad}$, $\rho_{\phi}$, $\rho_{\rm GB}$ and 
the total density $\rho$ as a function of the temperature $T$, 
for $\tilde{\alpha} = \pm 1$ $\rm km^2$, $\gamma = \pm 1$ and 
$\rho_\phi(T_{\rm BBN}) = 3\times10^{-2}\rho_{\rm BBN}$.}
\label{fig:rhos_gamma1}
\end{figure*}
\begin{figure*}[htb!]
\centering
\includegraphics[width=7.49cm,height=6.1cm]{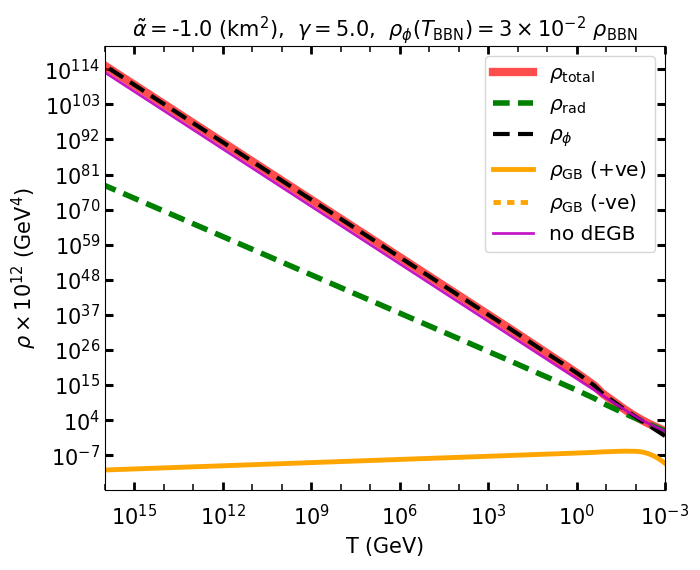}
\includegraphics[width=7.49cm,height=6.1cm]{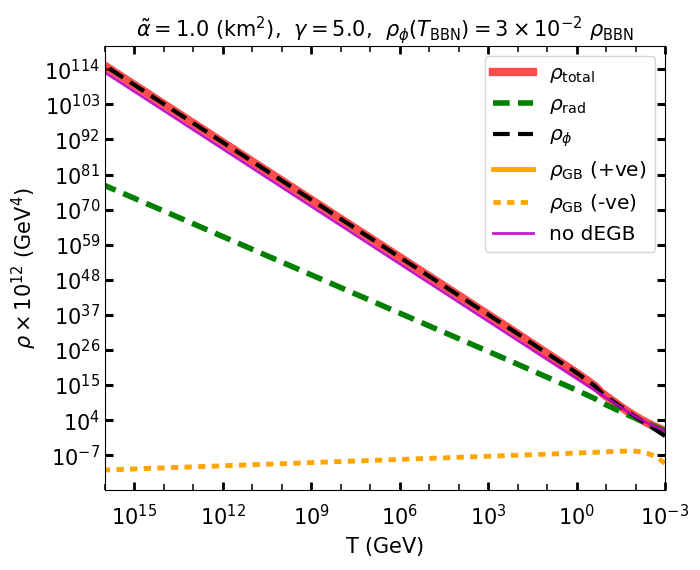}
\includegraphics[width=7.49cm,height=6.1cm]{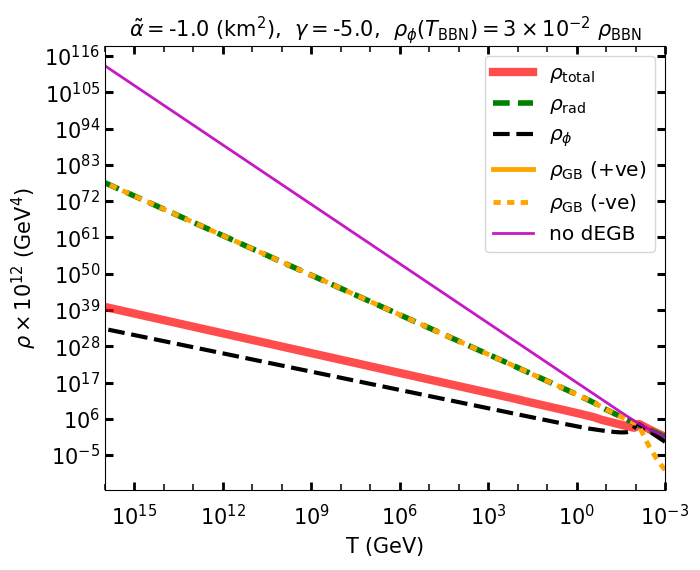}
\includegraphics[width=7.49cm,height=6.1cm]{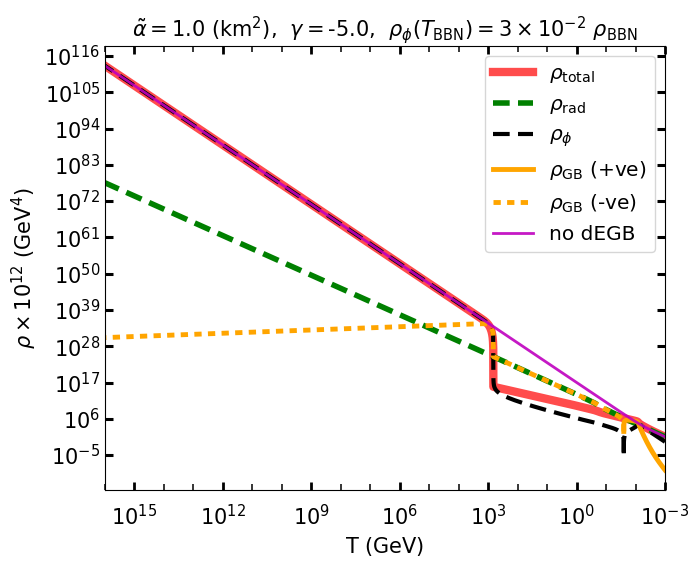}
\caption{Same as Fig.~\ref{fig:rhos_gamma1}, but for $\gamma = \pm 5$.}
\label{fig:rhos_gamma5}
\end{figure*}
\begin{figure*}[htb!]
\centering
\includegraphics[width=7.49cm,height=6.3cm]{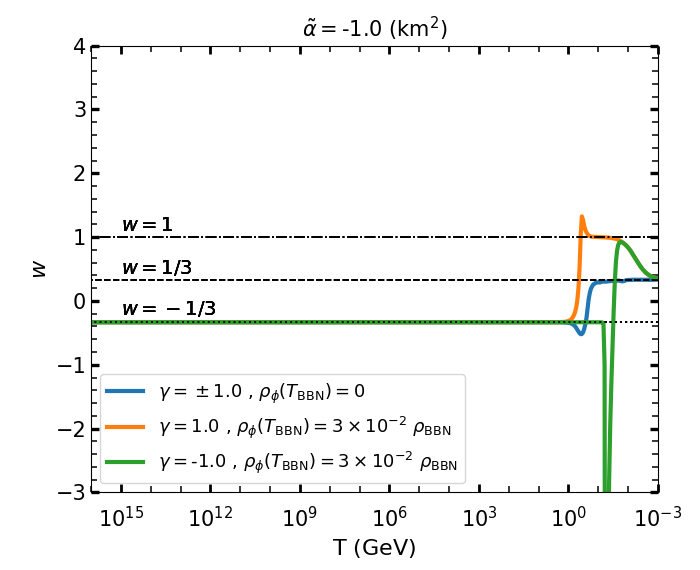}
\includegraphics[width=7.49cm,height=6.3cm]{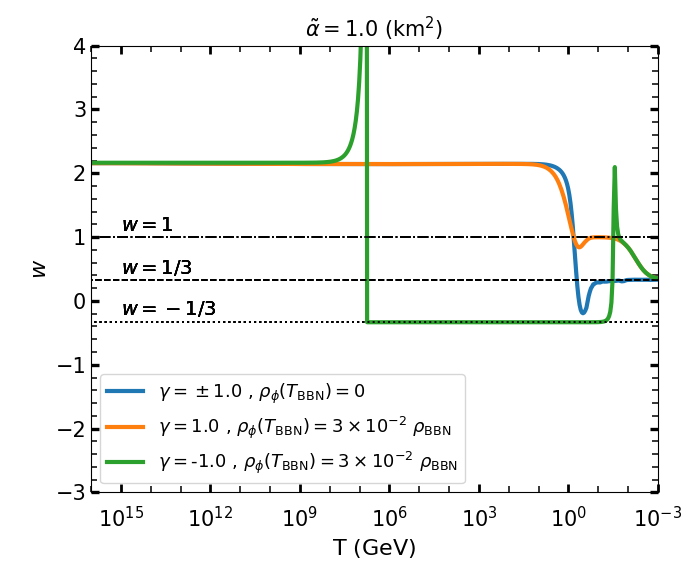}
\includegraphics[width=7.49cm,height=6.3cm]{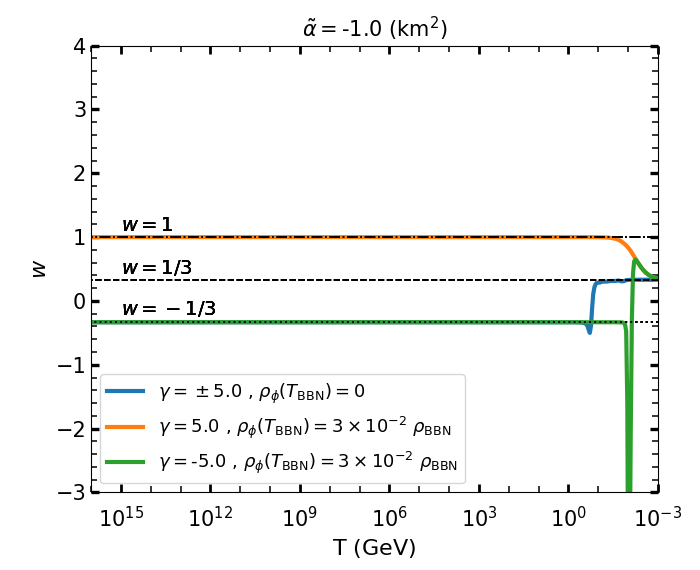} 
\includegraphics[width=7.49cm,height=6.3cm]{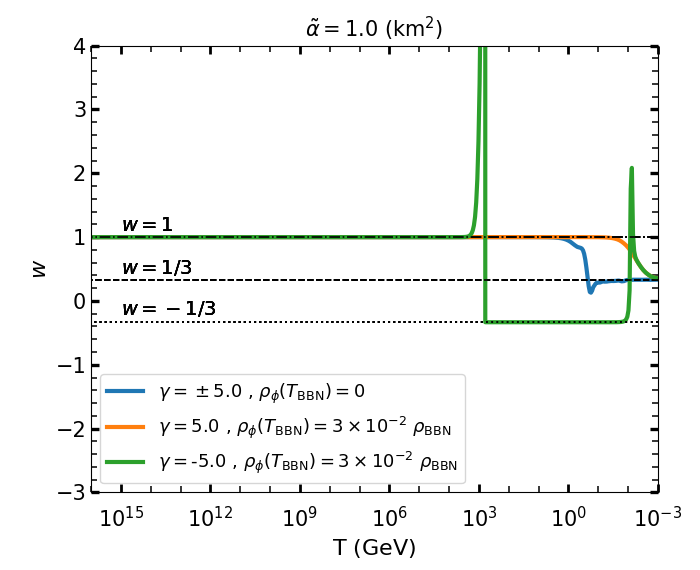}
\caption{Evolution of the equation of state of the Universe $w$ with $T$ 
for the same parameters of Figs.~\ref{fig:rhos_gamma1} and~\ref{fig:rhos_gamma5}.} 
\label{fig:eos}
\end{figure*}
\begin{figure*}[htb!]
\centering
\includegraphics[width=7.49cm,height=6.2cm]{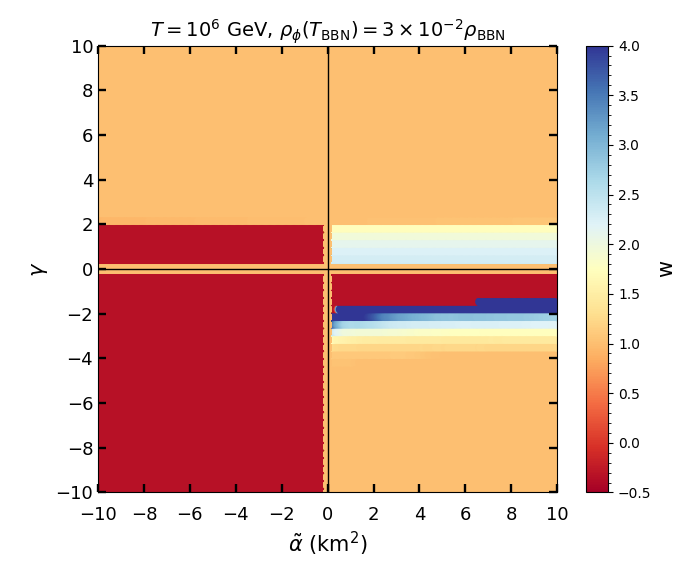}
\includegraphics[width=7.49cm,height=6.2cm]{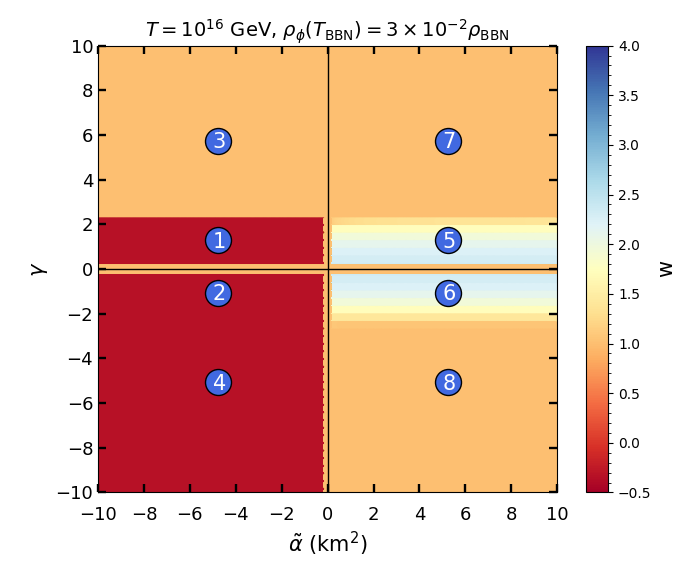}
\includegraphics[width=7.49cm,height=6.2cm]{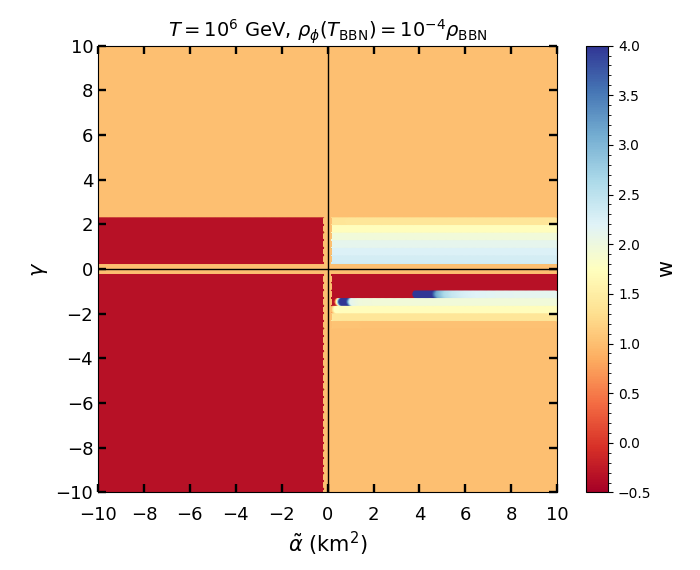}
\includegraphics[width=7.49cm,height=6.2cm]{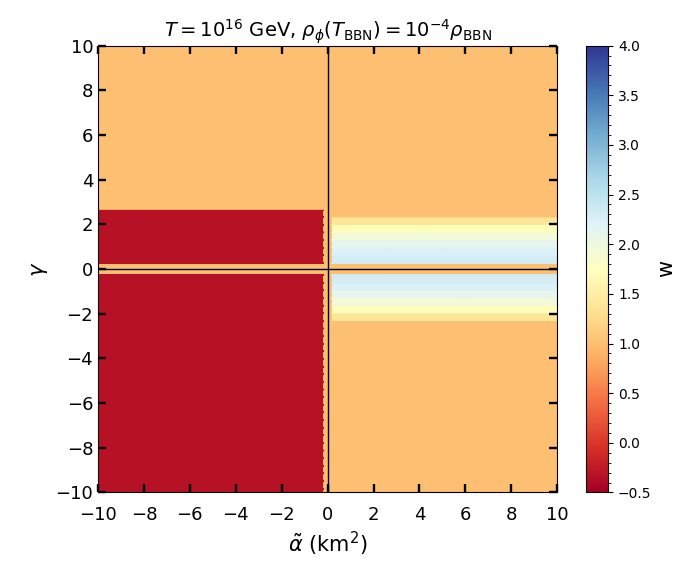}
\includegraphics[width=7.49cm,height=6.2cm]{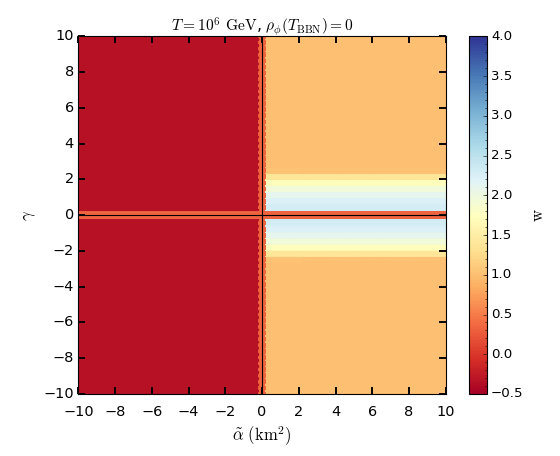}
\includegraphics[width=7.49cm,height=6.2cm]{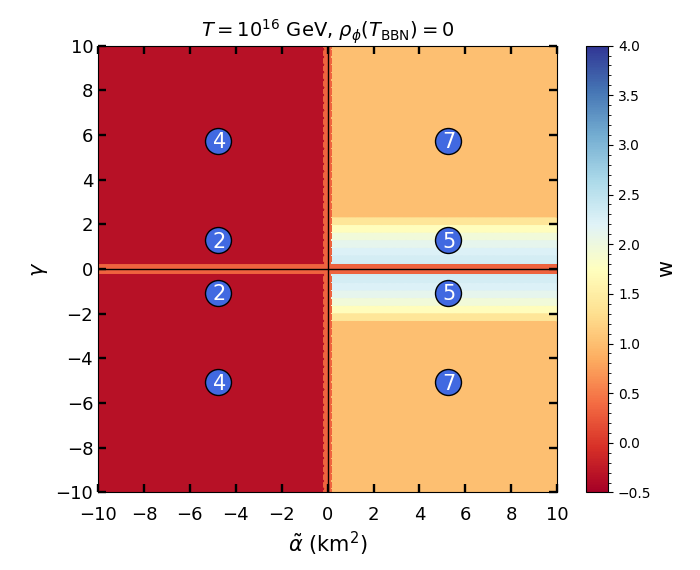}
\caption{The color codes show the variations of the equation of state ($w$) 
of the Universe in the 
Gauss-Bonnet parameter space (spanned by $\tilde{\alpha}$ and $\gamma$). 
{\it Top panels:} $\rho_\phi(T_{\rm BBN}) = 3\times10^{-2}\rho_{\rm BBN}$, 
{\it middle panels:} $\rho_\phi(T_{\rm BBN}) = 10^{-4}\rho_{\rm BBN}$, 
{\it bottom panels:} $\rho_\phi(T_{\rm BBN}) = 0$. {\it Left column:} $T=10^6$ GeV, 
{\it right column:} $T=10^{16}$ GeV.}
\label{fig:eos_alpha_gamma}
\end{figure*}

In Figs.~\ref{fig:rhos_gamma1} and \ref{fig:rhos_gamma5} we show a couple of examples for the numerical solutions of the differential equations. In particular, both figures  show the evolution of $\rho_{\rm rad}$, $\rho_{\phi}$, $|\rho_{\rm GB}|$ and $\rho$ as a function of the temperature $T$ when $\rho_{\phi}(T_{\rm BBN}) = 3\times10^{-2}\rho_{\rm BBN}$ and $\tilde{\alpha}=\pm 1$km$^2$. Moreover, in Fig.~\ref{fig:rhos_gamma1}  
 $\gamma = \pm 1$, 
while in Fig.~\ref{fig:rhos_gamma5} 
 $\gamma = \pm 5$.
As already pointed out, all our results are presented in terms of $\tilde{\alpha}$, i.e. we adopt the gauge $\phi(T_{\rm BBN}) = 0$. Inspection of such figures reveals a behaviour already observed in the analysis of Ref.~\cite{GB_WIMPS_sogang}: after a transient regime at low temperatures that can also show a high degree of nonlinearity, at large enough temperatures the energy density $\rho$ always reaches an asymptotic regime characterized by a constant equation of state $w=p/\rho$. This is shown in more detail in Fig.~\ref{fig:eos}, where in the upper panel the orange and green solid lines correspond to the evolution of $w$ for the same parameters of Fig.~\ref{fig:rhos_gamma1}, while the evolution of  $w$ shown in the lower panel corresponds to the parameters of Fig.~\ref{fig:rhos_gamma5}. Moreover, in Fig.~\ref{fig:eos} the blue solid lines include also the equation of state evolution for $\rho_{\phi}(T_{\rm{BBN}})=0$. In all such plots, most of the times $w$ appears to reach either $w = - 1/3$ or $w = 1$, with the exception of the cases $\tilde{\alpha}=+1 \, \rm{km}^2$, $\gamma = \pm 1$, for which $w\simeq 2.2$. In order to explore further this effect, a systematic evaluation of the asymptotic value of $w$ at high temperatures is provided in Fig.~\ref{fig:eos_alpha_gamma}, where the color code in each plot represents the value of $w$ in the $\tilde{\alpha}$ -- $\gamma$ plane. In particular, the plots in the left--hand column show $w$ at $T$ = $10^6$ GeV, while those in the right--hand one show the same quantity for  $T$ = $10^{16}$ GeV. Moreover, the plots in the first row are calculated for $\rho_{\phi}(T_{\rm BBN})=3\times 10^{-2}\rho_{\rm BBN}$, those in the central row for  $\rho_{\phi}(T_{\rm BBN})= 10^{-4}\rho_{\rm BBN}$ and those in the bottom row for $\rho_{\phi}(T_{\rm BBN})=0$. 

In such plots, two effects can be clearly seen when $T$ is large enough (and in particular in the case $T=10^{16}$ GeV, shown in the right--hand column of Fig. \ref{fig:eos_alpha_gamma}): i) for a wide range of the $\tilde{\alpha}$ and $\gamma$ parameters and for different values of $\rho_{\phi}(T_{\rm BBN})$, at large temperatures the system only saturates the two values $w$ =1 and $w = -1/3$, with the exception of the small interval $-2.45\lesssim\gamma\lesssim2.45$ for $\tilde{\alpha}>0$, where $w$ varies continuously in the range $1\lesssim w\lesssim 2.3$; ii) the boundaries of the different $w$ domains are flat in the $\tilde{\alpha}$ parameter, i.e. the asymptotic value of $w$ depends only on the sign of $\tilde{\alpha}$, but not on its actual value.

 In the numerical analysis of Ref.~\cite{GB_WIMPS_sogang}, the asymptotic behaviours summarized above were explained in terms of Eq.~(\ref{Eqfrw3}) for the scalar field $\phi$, and specifically by how the $V'_{GB}$ term affects the evolution of $\dot{\phi}$ at high $T$. A peculiar feature of such evolution is that, with the exception of a case when it is negligibly small, $\rho_{\rm GB}$ turns out to be negative at high $T$, no matter of its sign at $T_{BBN}$. 
 In particular, we note the following salient points: 
 \begin{itemize}
     \item In the parameter space where $w = 1$ one observes that $V'_{GB}$ and $\rho_{\rm GB}$ are exponentially suppressed and get quickly out of the game at high $T$, so that a standard kination scenario where $\rho$ is driven by the kinetic energy of the scalar field is recovered (i.e. a quintessence model~\cite{Tsujikawa:2013fta, Harko:2013gha, Cicoli:2018kdo, Bahamonde:2017ize, Alexander:2019rsc,  Banerjee:2020xcn} with vanishing potential $V(\phi)$).
     \item In the parameter space where $w = - 1/3$, the $V'_{GB}$ term mitigates and eventually stops the evolution of the scalar field at high $T$, so that asymptotically $\rho_{\phi}/\rho \rightarrow 0$, and $\rho\simeq \rho_{\rm rad}+\rho_{\rm GB}$ (we designate such regime, characterized by a subdominant $\dot{\phi}$, as ``slow roll"). In such condition,  $|\rho_{\rm GB}|$ eventually tracks $\rho_{\rm rad}$ closely, because it grows much faster than it due to its $H^3$ dependence and cannot exceed it to preserve the positivity of $\rho$, so $|\rho_{\rm GB}|$, $|\rho_{\rm GB}|\propto T^4$. However, the sum of the two terms has a lower $T$ dependence, $\rho\propto T^2$, due to the large cancellation between them\footnote{This peculiar cancellation will be instrumental in enhancing the GW stochastic background discussed in Section~\ref{sec:GW_constraints}.}.
     \item Finally, when $1 < w \lesssim$ 2.3 the $V'_{GB}$ term accelerates the evolution of $\phi$ beyond standard kination at high $T$ (we designate such regime as ``fast roll"), so that $\rho\simeq \rho_{\phi}+\rho_{\rm GB}$. Also, in this case $|\rho_{\rm GB}|$ tracks $\rho_{\phi}$, but the level of cancellation is not large unless $|\gamma| \rightarrow 0$.
 \end{itemize}

A more detailed discussion of what summarized above can be found in Ref.~\cite{GB_WIMPS_sogang}. In the next Section an appropriate change of variables will allow to explain all its peculiar features in a transparent way, by reformulating Eqs.~(\ref{Eqfrw1}, \ref{Eqfrw2}) in terms of a system of autonomous differential equations with a simple pattern of critical points and attractors.

\begin{figure*}[htb!]
\centering
\includegraphics[width=7.49cm,height=6.5cm]{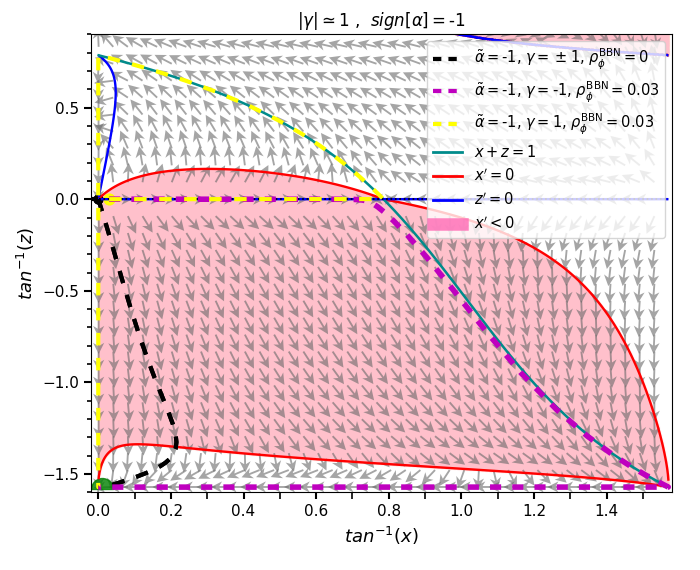}
\includegraphics[width=7.49cm,height=6.5cm]{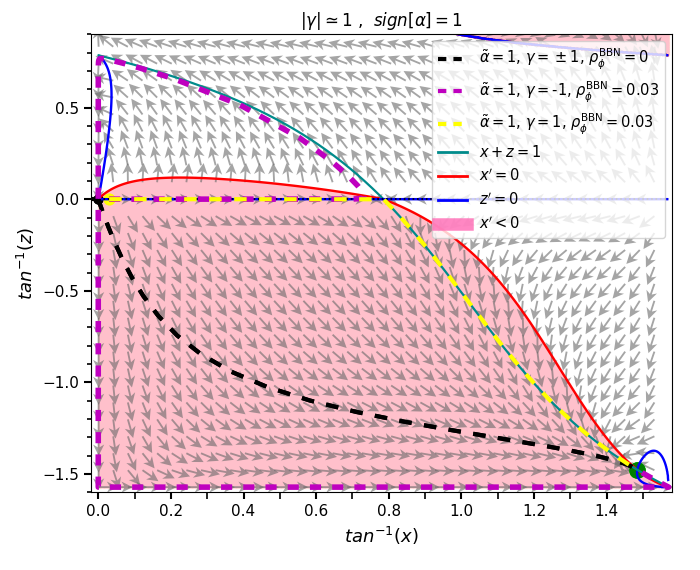}
\includegraphics[width=7.49cm,height=6.5cm]{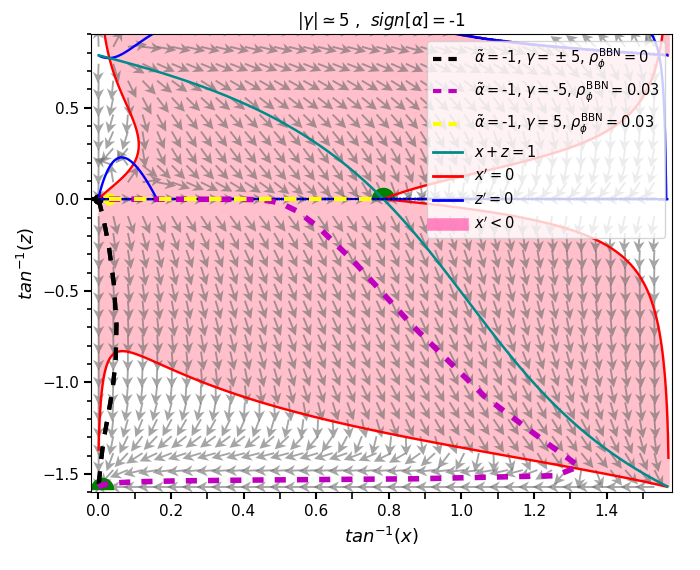}
\includegraphics[width=7.49cm,height=6.5cm]{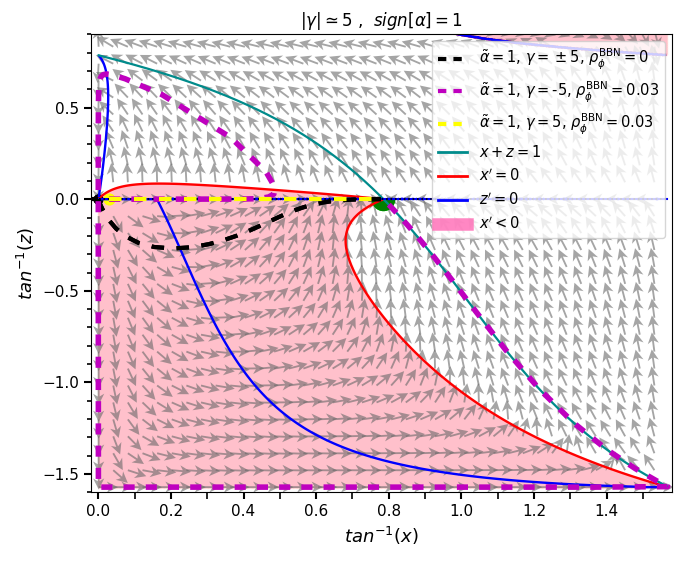}
\caption{Examples of the evolution of Eqs~(\ref{eq:x_prime}, \ref{eq:z_prime}) in the $(x,z)$ plane (in the captions $\kappa$ =1). The gray arrows in the background represent the velocity field $(x^\prime,z^\prime)$, while the red and the blue solid lines correspond to $x' = 0$ and $z' = 0$, respectively. The red shaded areas indicate $x' = \frac{dx}{dN} < 0$. The solid cyan line corresponds to $x+z=1$ and the parameter space below this line is the physical one. In each case, the attractor point(s) is indicated in green. }
\label{fig:flow_solution}
\end{figure*}

\section{Asymptotic behaviours}
\label{sec:tracking}

\subsection{Equivalent system of autonomous equations}
\label{sec:autonomous}

We start this Section by introducing the following variables~\cite{Koh:2014fxg, Myung:2015tga, Odintsov:2017tbc, Chatzarakis:2019fbn}:
\begin{eqnarray}
x &\equiv& \frac{\rho_\phi}{\rho}=\frac{\kappa}{6} \left(\frac{{\dot \phi }}{H}\right)^2\,, \nonumber\\
y &\equiv& \frac{\rho_{\rm rad}}{\rho}=\frac{\kappa g_*\pi^2}{90} \frac{T^4}{H^2} \,, \nonumber\\
z &\equiv& \frac{\rho_{\rm GB}}{\rho}=-8\kappa {\dot f} H=-8\kappa \frac{\partial f}{\partial \phi}\dot{\phi}H.
\label{eq:def_xyz}
\end{eqnarray}

\noindent In terms of $x,y,z$ the system of coupled differential equations (\ref{Eqfrw2}, \ref{Eqfrw3}) can then be re--expressed in the following form:

\begin{eqnarray}
x' &=&2x\left(\frac{\beta}{z}+\epsilon-\mu\sqrt{x} \right)=2\left (\epsilon-3\right ) x+\left (\epsilon-1\right )z,\label{eq:xprime}\nonumber\\
y' &=& 2\left (\epsilon -2\right ) y,\label{eq:yprime}\nonumber\\
z' &=& \beta-\epsilon z=6x+4y+\left (1+\epsilon\right )z-2\epsilon,
\label{eq:diff_xyz}
\end{eqnarray}
\noindent where the prime indicates a derivative with respect to the e--folding parameter $N=\log(a)$ ($'=\frac{d}{dN}=\frac{d}{d \ln a}=\frac{1}{H}\frac{d}{dt}$) and

\begin{eqnarray}
    \beta&\equiv& -8\kappa \ddot{f},\label{eq:beta}\\
    \epsilon&\equiv& -\frac{\dot{H}}{H^2}=1+q,
    \label{eq:epsilon}
\end{eqnarray}
\noindent where $q=-\ddot{a}a/\dot{a}^2$ is the deceleration parameter. Moreover, we define 

\begin{equation}
\mu\equiv \sqrt{\frac{\kappa}{6}}\frac{\partial^2 f}{\partial \phi^2}/\frac{\partial f}{\partial \phi}=\sqrt{\frac{6}{\kappa}}\gamma,   
\end{equation}
\noindent where in the last equality the explicit expression of $f$ in~(\ref{eq:f_GB}) is used. In Eqs.~(\ref{eq:def_xyz}) $x$ and $y$ are positive defined, while $z$ can have both signs.  Furthermore $x$, $y$ and $z$  verify the identities:
\begin{eqnarray}
x+y+z&=&1,\label{eq:xyz_sum}\\
x^{\prime}+y^{\prime}+z^{\prime}&=&0,
\label{eq:xyz_prime_sum}
\end{eqnarray}

\noindent and the system contains only two independent degrees of freedom. Choosing them to be $x$ and $z$ Eqs.~(\ref{eq:diff_xyz}) yield
\begin{eqnarray}
    x'&=& 2\left [\epsilon(x,z)-3\right ]x +\left[\epsilon(x,z)-1\right]z \equiv F(x,z)\label{eq:x_prime},\\
    z'&=& 2 x +\left[\epsilon(x,z)-3\right]z +2\left[2-\epsilon(x,z)\right] \equiv G(x,z),
    \label{eq:z_prime}
\end{eqnarray}

\noindent where $\epsilon(x,z)$ is explicitly given by
\begin{eqnarray}
    \epsilon(x,z)&=&\frac{4x^2+8x+z^2+2\sqrt{\frac{6}{\kappa}}{\rm sign}(\alpha)|\gamma||z|x^{3/2}}{4x-4xz+z^2}.\label{eq:epsilon_xz} 
\end{eqnarray}
\noindent Note that in the equation above the appearance of the term ${\rm sign}(\alpha)|\gamma||z|$ = ${\rm sign}(\alpha \gamma z) \gamma z$ is due to the fact that the change of variables $(\phi,\dot{\phi})\rightarrow (x,z)$  requires to keep track of the sign of $\dot{\phi}$, since $z\propto \rho_{\rm GB}$ is linear in it. Explicitly: $\dot{\phi}= {\rm sign}(\dot{\phi}) \sqrt{2\rho_{\phi}}$ = ${\rm sign}(\dot{\phi})H\sqrt{\frac{6}{\kappa}} x^{1/2}$ = $-{\rm sign}(\alpha\gamma z)H\sqrt{\frac{6}{\kappa}} x^{1/2}$, with ${\rm sign}(\dot{\phi})$ = $-{\rm sign}(\alpha\gamma\rho_{\rm GB})$ = $-{\rm sign}(\alpha\gamma z)$. 

The two coupled non--linear differential equations (\ref{eq:x_prime},  \ref{eq:z_prime}) are autonomous~\cite{Coddington_Levinson_1955}, i.e. they do not depend explicitly on $N$ and describe a velocity field $(x',z')$ in the $x$ -- $z$ plane. An explicit example of the evolution of Eqs.~(\ref{eq:x_prime}, \ref{eq:z_prime}) is provided in Fig.~\ref{fig:flow_solution}. In such figure, the velocity field  is represented by the pattern of gray arrows in the background. The asymptotic behaviours at large $T$ discussed in the previous Section correspond to its attractors at $N\rightarrow -\infty$, i.e. to the $(x_c, z_c)$ configurations in the $(x, z)$ plane for which  $x'=0,z'=0$ (critical points) which are also stable\footnote{The stability of the critical points depends on the direction of the time variable in which they are solved. In our analysis we will always refer to attractors, repellers and saddle points in the case when the differential equations~(\ref{eq:x_prime},\ref{eq:z_prime}) are evolved from $T_{BBN}$ to higher temperatures. This convention also fixes the direction of the velocity field shown in Fig.~\ref{fig:flow_solution} and of the evolution patterns of Fig.~\ref{fig:critical_points}.}.  In particular, if $(x_c,z_c)$ is an attractor $\epsilon(x,z)$ saturates to the constant value $\epsilon(x_c,z_c)$, and this fixes the equation of state $w$, since the two quantities are related by
\begin{equation}
    w=\frac{2}{3} \epsilon -1.
    \label{eq:w_epsilon}
\end{equation}
\noindent So attractors explain the peculiar behaviour observed in the numerical plots of Figs.~\ref{fig:eos}, where $w$ saturates to a constant value at high $T$. Moreover, through $\epsilon(x,z)$ (see Eq.~(\ref{eq:epsilon_xz})) the set of autonomous differential equations~(\ref{eq:x_prime}, \ref{eq:z_prime}), and so its critical points, depend  only on the sign of $\alpha$ but not on its actual value. This explains why at large--enough temperatures the equation of state domains of Fig.~\ref{fig:eos_alpha_gamma} appear as horizontal bands in the $\tilde{\alpha}$--$\gamma$ plane. In the following section we will find the critical points of Eqs.~(\ref{eq:x_prime}, \ref{eq:z_prime}) and discuss if/when they are stable attractors. 

\subsection{Finite critical points}
\label{sec:finite_critical_points}

Inverting Eqs.~(\ref{eq:x_prime}, \ref{eq:z_prime}) with $x'=0, z'=0$, we obtain the following implicit form:
\begin{eqnarray}
    x_c&=&\frac{\epsilon - 1}{5-\epsilon},\nonumber\\
    z_c&=&-2\frac{\epsilon -3}{5-\epsilon}\, ,
    \label{eq:xz_critical_points}
\end{eqnarray}
\noindent for the critical points. Substituting back $x_c$ and $z_c$ in (\ref{eq:epsilon_xz}) we obtain the following equation for $\epsilon$:
\begin{equation}
    (\epsilon -1)(\epsilon -3)\left( \sqrt{\frac{6}{\kappa}}|\gamma| {\rm sgn}(\alpha z_c) \sqrt{\frac{\epsilon-1}{5-\epsilon}}+2\epsilon \right) =0,
    \label{eq:epsilon_xy}
\end{equation}
\noindent which has three different solutions. The first two critical points are always present and do not depend on the parameters $(\alpha,\gamma)$. Specifically:  
\begin{itemize}
    \item $\epsilon=3\rightarrow (x_c,z_c)=(1,0)$\;\;\mbox{(kination)};
    \item $\epsilon=1\rightarrow(x_c,z_c)=(0,1)$\;\;\mbox{(GB)}.
\end{itemize}
\noindent In the following, we will refer to them as the ``kination" and the ``GB" critical point (see also Table~\ref{table:critical_points}). On the other hand, the third critical point, that we will indicate with $(x_{\rm fp},z_{\rm fp})$, is obtained by setting to zero the expression in the third parenthesis of Eq.~(\ref{eq:epsilon_xy}), which vanishes only if ${\rm sign}(\alpha z_{\rm fp})<$0. If the latter condition is verified $x_{\rm fp}$ and $z_{\rm fp}$ are given by Eq.~(\ref{eq:xz_critical_points}) with $\epsilon$ being the only real solution of the cubic equation\footnote{A cubic equation $ax^3+bx^2+cx+d$ has one real and two complex solutions when its discriminant $\Delta=18abcd-4b^3d+b^2c^2-4ac^3-27a^2d^2$ is negative. The discriminant of Eq.~(\ref{eq:cubic}) (with $\kappa=1$) is $\Delta=-8\gamma^2(-27\gamma^4+396\gamma^2-1500)$ and is negative for any value of $\gamma\ne0$.}

\begin{equation}
    2 \epsilon^3-10 \epsilon^2+\frac{3}{\kappa}\gamma^2\epsilon-\frac{3}{\kappa}\gamma^2=0.
    \label{eq:cubic}
\end{equation}

\begin{figure*}[htb!]
\centering
\includegraphics[width=7.49cm,height=6cm]{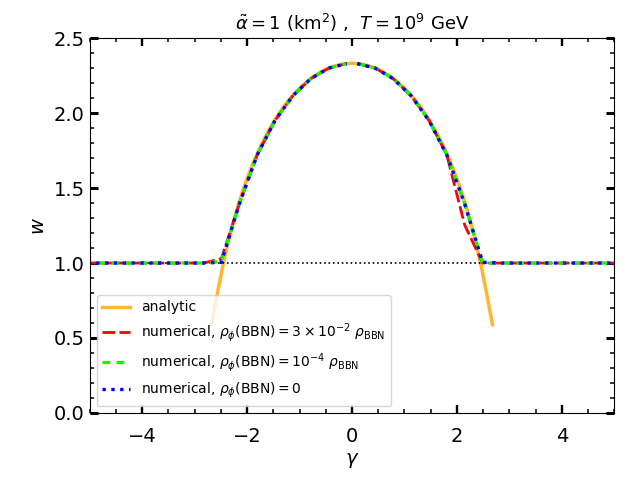}
\includegraphics[width=7.49cm,height=6cm]{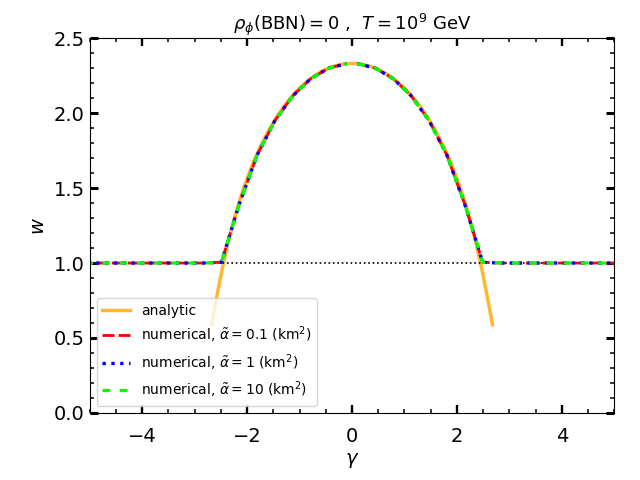}
\caption{Equation of state $w$ for the critical point $(x_{\rm fp},z_{\rm fp})$ as a function of $\gamma$ and for $\alpha>0$. In the caption $\kappa = 1$. The yellow solid line represents the equation of state $w$ calculated using the solution of the cubic equation~(\ref{eq:cubic}). The other curves correspond to the numerical solution of the full set of Eqs.~(\ref{Eqfrw1}, \ref{Eqfrw2}, \ref{Eqfrw3}). {\it Left:} for different values of $\rho_\phi(T_{\rm BBN})$; 
{\it Right:} for different values of $\tilde{\alpha}$.}
\label{fig:w_gamma_fastRoll_analytical}
\end{figure*}

\noindent  In both plots of Fig.~\ref{fig:w_gamma_fastRoll_analytical} the orange solid line shows the equation of state $w=2/3\epsilon-1$ as a function of $\gamma$, with $\epsilon$ solution of Eq.~(\ref{eq:cubic}).
In particular Eq.~(\ref{eq:cubic}) is equivalent to the condition $\sqrt{6/\kappa}|\gamma|\sqrt{x_{\rm fp}}=2\epsilon$, that when $w=1$,  $\epsilon=3$ and $x_{\rm fp}=1$ yields $\gamma=\sqrt{6\kappa}$. Fig.~\ref{fig:w_gamma_fastRoll_analytical} shows that $|\gamma|<\sqrt{6\kappa}$ when $\epsilon>3$ and $|\gamma|>\sqrt{6\kappa}$ when $\epsilon<3$. Then the requirement ${\rm sign}(\alpha z_{\rm fp})<$0 implies two possibilities:
\begin{itemize}
    \item $\alpha>0$. In this case $z_{\rm fp}<0$ and $\epsilon>3$, which requires that in Eq.~(\ref{eq:cubic}) $|\gamma|<\sqrt{6\kappa}$;
    \item $\alpha<0$. In this case $z_{\rm fp}>0$ and $\epsilon<3$, which requires that in Eq.~(\ref{eq:cubic}) $|\gamma|>\sqrt{6\kappa}$.
\end{itemize}

The critical points in the $x$--$z$ plane are shown in a pictorial way in
Fig.~\ref{fig:critical_points}, where the 4 possible sign combinations of $\alpha$ and $|\gamma|-\sqrt{6\kappa}$ are separately shown. In such figure the three critical points $(1,0)$, $(0,1)$ and $(x_{\rm fp},z_{\rm fp})$ are represented by circles filled in green or in red color if they are stable or unstable, respectively, according to the discussion below. In particular, the $(x_{\rm fp}, z_{\rm fp})$ critical point lies on the $x+z=1$ line and varies in the interval $(+\infty,-\infty)\rightarrow(1,0)$ when $0<|\gamma|<\sqrt{6\kappa}$ and $\alpha>0$ , while it varies in the interval $(1,0)\rightarrow (0,1)$ when $\sqrt{6\kappa}<|\gamma|<\infty$ and for $\alpha<0$. 

In order to determine whether a critical point $X_c=(x_c, z_c)$ is stable, one needs to expand up to linear order the velocity field $X'=(x',z')$ in its neighborhood

\begin{eqnarray}
    x^\prime&=&F(x,z)\simeq F(x_{c},z_{c})+\frac{\partial F}{\partial x}(x-x_c)+ \frac{\partial F}{\partial z}(z-z_c),\nonumber\\
    z^\prime&=&G(x,z) \simeq G(x_{c},z_{c})+\frac{\partial G}{\partial x}(x-x_c)+ \frac{\partial G}{\partial z}(z-z_c),
    \label{eq:xz_general}
\end{eqnarray}

\noindent which, using  $F(x_{c},z_{c})$ = $G(x_{c},z_{c})$ = 0, is equivalent to

\begin{equation}
    X'=J (X-X_c),
    \label{eq:critical_point_expansion}  
\end{equation}
 
\noindent where

\begin{equation}
    J=
 \left(\begin{array}{cc}
     \frac{\partial F}{\partial x} & \frac{\partial F}{\partial z} \\
     \frac{\partial G}{\partial x} & \frac{\partial G}{\partial z} 
        \end{array}\right )_{(x,z)=(x_c,z_c)}.
    \label{eq:xz_MatrixGeneral}
\end{equation}

\noindent $X_c$ is a stable attractor for $N\rightarrow -\infty$ if the eigenvalues of the Jacobian J are both positive. It is explicitly given by
\begin{equation}
J= \left(\begin{array}{cc} 2\left [\epsilon(x_c,z_c)-3\right ] & \epsilon(x_c,z_c)-1 \\
                  2 & \epsilon(x_c,z_c)-3 
         \end{array}\right )+(x_c+1)\left(\begin{array}{cc}
\frac{\partial \epsilon}{\partial x} & \frac{\partial \epsilon}{\partial z} \\
-\frac{\partial \epsilon}{\partial x} & 
-\frac{\partial \epsilon}{\partial z}
\end{array}\right )_{(x,z)=(x_c,z_c)},
\label{eq:J_matrix}
\end{equation}

\noindent where
\begin{eqnarray}
 \frac{\partial\epsilon}{\partial x} &=&\frac{3\sqrt{\frac{6}{\kappa}}{\rm sgn}{(\alpha z)}|\gamma|  x^{1/2}z+8x+8+4(z-1)\epsilon}{4x-4xz+z^2},\\
\frac{\partial\epsilon}{\partial z} &=&\frac{2\sqrt{\frac{6}{\kappa}}{\rm sgn}{(\alpha z)}|\gamma|x^{3/2} +2z+(4x-2z)\epsilon}{4x-4xz+z^2}.
    \label{del_epsilon_over_del_xz}
\end{eqnarray}

In particular, substituting $(x_c,z_c)$ = $(0,1)$ in Eq.~(\ref{eq:J_matrix}) $J$ yields

\begin{eqnarray}
J&=& \left(\begin{array}{cc} 4 & 0  \nonumber \\
       -6 & -2 
         \end{array}\right ).\nonumber \\
{\rm Eigenvalues:}\quad \lambda_1&=&4, \quad \lambda_2=-2, \nonumber\\
{\rm Eigenstates:}\quad v_1&=&[1/\sqrt{2},-1/\sqrt{2}], \quad v_2=[0,1],
\label{eq:eigen_GB}
\end{eqnarray}

\noindent so that the critical point $(0,1)$ is a saddle point and is not stable. For this reason, in Fig.~\ref{fig:critical_points} the corresponding circle is always filled in red color\footnote{One should notice that  the critical point $(0,1)$ is not physical, because $x=0$ implies $\dot{\phi}$ = 0 and so also $z$ = $\rho_{\rm GB}$ = 0.}.  On the other hand, substituting $(x_c,z_c)$ = $(1,0)$ one gets

\begin{eqnarray}
&&J= \left(\begin{array}{cc} 2 & {\rm sign}(\alpha z)\sqrt{\frac{6}{\kappa}}|\gamma|+8  \nonumber\\
       0 &  -{\rm sign}(\alpha z)\sqrt{\frac{6}{\kappa}}|\gamma|-6
         \end{array}\right ),\nonumber\\
&&{\rm Eigenvalues:}\, \lambda_1=2,\,\,\lambda_2=-6-{\rm sign}(\alpha z)\sqrt{\frac{6}{\kappa}}|\gamma|,\nonumber\\
&&{\rm Eigenstates:}\, v_1=[1,0]\,\,v_2=[0,1].
\label{eq:GB_eigenvector}
\end{eqnarray}

In this case $(1,0)$
is an attractor (i.e. both eigenvalues are positive)  if ${\rm sign}(\alpha z)|\gamma|<-\sqrt{6\kappa}$, i.e. if  ${\rm sign}(\alpha z)<0$ {\it and} $|\gamma|>\sqrt{6\kappa}$. In other words:

\begin{eqnarray}
|\gamma|<\sqrt{6\kappa} &\rightarrow& \mbox{saddle point}, \nonumber\\
|\gamma|>\sqrt{6\kappa} &\rightarrow& \left\{\begin{array}{cc} \alpha>0 & \left\{\begin{array}{cc} z\rightarrow 0^+ & \mbox{saddle point}  \\
z\rightarrow 0^- & \mbox{attractor}  \end{array}\right . \\
\alpha<0 & \left\{\begin{array}{cc} z\rightarrow 0^+ & \mbox{attractor}  \\
z\rightarrow 0^- & \mbox{saddle point} . \end{array}\right .
         \end{array}\right .
\end{eqnarray}
\noindent As a consequence, in  Fig.~\ref{fig:critical_points} the circle corresponding to the critical point $(1,0)$ is filled in red (saddle point) when $|\gamma|<\sqrt{6\kappa}$ while it is filled in red/green in its positive/negative half depending on the sign of $\alpha$.

\begin{figure*}[htb!]
\centering
\includegraphics[width=7.49cm,height=6cm]{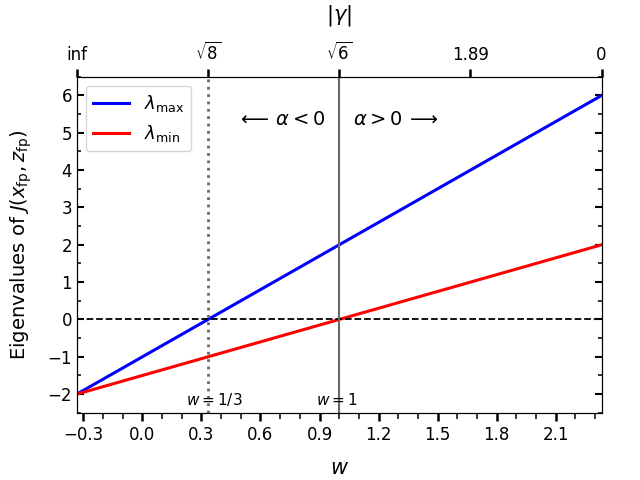}
\includegraphics[width=7.49cm,height=6cm]{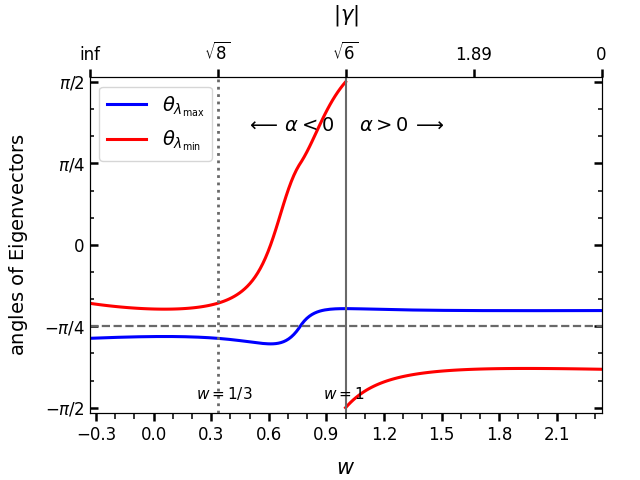}
\caption{Eigenvalues of the Jacobian $J(x_{\rm fp},z_{\rm fp})$ for the critical point $(x_{\rm fp},z_{\rm fp})$, defined by Eq.~(\ref{eq:xz_critical_points}) with $\epsilon$ solution of the cubic equation~(\ref{eq:cubic}) (in the plot $\kappa$ =1). The vertical dotted line corresponds to $|\gamma|$ = $\sqrt{8\kappa}\simeq2.83$, $\epsilon$ = 2 and $w=1/3$, while the vertical solid line corresponds to $|\gamma|$ = $\sqrt{6\kappa}\simeq2.45$, $\epsilon$ = 3 and $w=1$. Notice that $(x_{\rm fp},z_{\rm fp})$ is a critical point only for $\gamma<\sqrt{6\kappa}$  when $\alpha>0$, and for $\gamma>\sqrt{6\kappa}$  when $\alpha<0$ (see schematic view of Fig.~\ref{fig:critical_points}). 
\label{fig:jacobian_eigenvalues_2}}
\end{figure*}

As far as the critical point $(x_{\rm fp},z_{fp})$
is concerned, the analytic expressions of the Jacobian $J$ eigenvalues are cumbersome, so we provide a plot of their numerical values in Fig.~\ref{fig:jacobian_eigenvalues_2}. From this figure one can see that the two eigenvalues of the Jacobian are both positive (and, as a consequence, the critical point $(x_{\rm fp},z_{fp})$ is a stable attractor) only when $\alpha>0$ and $|\gamma|<\sqrt{6\kappa}$. Specifically, in the interval $0\le|\gamma|\le\sqrt{6\kappa}$ Eq.~(\ref{eq:cubic}) yields $3\le\epsilon\le 5$, which using Eq.~(\ref{eq:w_epsilon}) corresponds to  the interval 1 $\le w\le 7/3$ for the equation of state. This explains the upper bound $w\lesssim$ 2.3 observed in the numerical analysis of Ref.~\cite{GB_WIMPS_sogang} and in Section~\ref{sec:numerical} for $\alpha>0$ and $|\gamma|\rightarrow0$. In Table~\ref{table:critical_points} we refer to this attractor as ``fast--roll". On the other hand, when  $|\gamma|>\sqrt{6\kappa}$ and $\alpha<0$ such critical point is not stable, because either one or both the eigenvalues of the Jacobian are negative. The situation is again summarized schematically in Fig.~\ref{fig:critical_points}, where $(x_{\rm fp},z_{fp})$ is represented with a green dot (attractor) for $\alpha>0$ and $|\gamma|<\sqrt{6\kappa}$, with a red dot for  $\alpha<0$ and $|\gamma|>\sqrt{6\kappa}$, while it is missing in the remaining cases when $(x_{\rm fp},z_{fp})$  is not a critical point in the first place (i.e. for $\alpha>0$ and $|\gamma|>\sqrt{6\kappa}$ or $\alpha<0$ and $|\gamma|<\sqrt{6\kappa}$).

Combining the discussion of the critical points $(1,0)$ and ($x_{\rm fp},z_{\rm fp}$) it is now possible to understand why in Fig.~\ref{fig:w_gamma_fastRoll_analytical} for $\alpha>0$ the equation of state of the numerical solutions of the full system of differential equations (\ref{Eqfrw1}, \ref{Eqfrw2}, \ref{Eqfrw3}) tracks the analytical solution of the cubic equation~(\ref{eq:cubic}) for $|\gamma|<\sqrt{6\kappa}$ , so that in such case $1<w<7/3$, while for $|\gamma|>\sqrt{6\kappa}$ it saturates to $w=1$. In fact, in the schematic summary of Fig.~\ref{fig:critical_points} one can see that when $\alpha>0$ and $|\gamma|<\sqrt{6\kappa}$ (tracks $\circled{5}$ and $\circled{6}$ in Figs.~\ref{fig:eos_alpha_gamma} and \ref{fig:critical_points}) the only stable critical point of the system is $(x_{\rm fp},z_{\rm fp})$, while for  $|\gamma|>\sqrt{6\kappa}$ it is $(1,0)$ (for $z\rightarrow 0^{-}$, see tracks $\circled{7}$ and $\circled{8}$ in the same figures).

As a comment one can also point out that from Fig.~\ref{fig:jacobian_eigenvalues_2} one can see that $(x_{\rm fp},z_{\rm fp})$ is a saddle point (with $\lambda_{\rm max}>0$ and $\lambda_{\min}<0$) for $\sqrt{6\kappa}<|\gamma|<\sqrt{8\kappa}$, corresponding to $1/3<w<1$, while it is  a repelling node (with both negative eigenvalues, $\lambda_{\rm max},\lambda_{\min}<0$) for $|\gamma|>\sqrt{8\kappa}$, corresponding to $-1/3<w<1/3$. 
In particular, when $(x_{\rm fp},z_{\rm fp})$ is a saddle point one of the two normal modes corresponds to an unstable attractor, that can be achieved by a tuning of the boundary conditions. However, such attractor is only possible when $w>1/3$, while for $w<1/3$ such unstable solution disappears. Why this transition happens for $w=1/3$, corresponding to the equation of state of radiation, can be understood in the following way. The right--hand plot of Fig.~\ref{fig:jacobian_eigenvalues_2} shows the angle between the approaching path to $(x_{\rm fp},z_{\rm fp})$ of each normal mode of the linearized differential equation (\ref{eq:critical_point_expansion}) and the $x$ axis. Such angle is never $-\pi/4$, i.e. the normal modes are always outside the path along the $x+z=1$ line. Formally, this means that the boundary condition to obtain both normal modes always requires $x+y<1$ and $y>0$. Physically, this means that such asymptotic cosmological solution requires a non--vanishing radiation density at lower temperatures.  This includes the case of the unstable attractor corresponding to $\lambda_{\rm max}>0$. Such a solution is only possible if the expansion rate of the Universe is faster than that of radiation, i.e. for  $w>1/3$. On the other hand, when $w<1/3$ radiation is bound to eventually dominate at high temperature, and, even at the cost of tuning, such $(x_{\rm fp},z_{\rm fp})$ solution is not achievable anymore. As a consequence, also the second eigenvalue in the left-hand plot of Fig.~\ref{fig:jacobian_eigenvalues_2} turns negative. 
This is at variance with what happens to the case of the $(0,1)$ saddle point discussed before, which has an asymptotic equation of state as low as $w=-1/3$, corresponding to an expansion rate slower than radiation. In this case, the direction of the unstable attractor that corresponds to the positive eigenvalue $\lambda$ = 4 is along the line $x+z$ =1 (see Eq.(\ref{eq:eigen_GB})) requiring, identically $y$ = $\rho_{\rm rad}$ =0. 
Physically, such cosmological solution at $T\rightarrow\infty$ requires that the radiation density exactly vanishes at some lower temperature so that the corresponding expansion rate $\rho\propto T^2$ is not overtaken at high temperature by that of radiation, $\rho_{\rm rad}\propto T^4$.

\subsection{Critical points at infinity}
\label{sec:attractors_at_infinity}

In the previous Section when $|\gamma|\rightarrow$ 0 Eq.~(\ref{eq:cubic}) yields $\epsilon\rightarrow$ 5 and in Eq.~(\ref{eq:xz_critical_points}) the critical point $(x_{\rm fp},z_{\rm fp})$ is driven to $(+\infty,-\infty)$ along the line $x+z=1$, with $w\rightarrow$ 7/3. In this case $\rho_{\phi}\rightarrow\infty$ and $\rho_{\rm GB}\rightarrow -\infty$, with a large cancellation between the two, so that the total density remains finite, while at the same time $\rho_{\rm rad}/\rho\rightarrow$0. This effect is possible because $\rho_{\rm GB}$ can grow negative at high temperature.

Something analogous happens when at high temperature a large cancellation occurs between $\rho_{\rm rad}\rightarrow\infty$ and $\rho_{\rm GB}\rightarrow -\infty$ while, at the same time, $\rho_{\phi}/\rho\rightarrow 0$. In order to see that this can be a stable regime at high temperature one can expand Eq.~(\ref{eq:epsilon_xz}) in the leading term ${\cal O}(x/z)\rightarrow$ 0, obtaining

\begin{eqnarray}
    &&\epsilon=\frac{1+{\cal O}(\frac{x^{3/2}}{z})}{1-4\frac{x}{z}+{\cal O}(\frac{x}{z^2})}\simeq 1+4\frac{x}{z}\nonumber\\\Rightarrow &&(\epsilon-1)z\rightarrow 4x,
    \label{eq:epsilon_slow_roll}
\end{eqnarray}
\noindent which implies $\epsilon\rightarrow$1 and $w\rightarrow -1/3$, while, at the same time, $x^{\prime}=0$ (when the equation above is used in~(\ref{eq:x_prime})). This slow--roll regime for the scalar field was already discussed in~\cite{GB_WIMPS_sogang}, and has the peculiarity that $\rho_{\rm rad},|\rho_{\rm GB}|\propto T^4$ but $\rho_{\rm rad}+\rho_{\rm GB}\propto T^{3(1+w)}\propto T^2$, with a large cancellation between $\rho_{\rm rad}$ and $\rho_{\rm GB}$. Indeed, combining $x^{\prime}=0$ with Eq.~(\ref{eq:xyz_prime_sum}) implies $(y+z)^{\prime}=0$, while at the same time for $\epsilon\rightarrow$1 one has $y^{\prime}=-2y\ne 0$ from Eq.~(\ref{eq:diff_xyz}). As a consequence both $y^{\prime}\simeq -z^{\prime}\ne0$ but $y^{\prime}+z^{\prime}=0$. In order to understand if such critical point at infinity is stable one needs to extend the expansion~(\ref{eq:epsilon_slow_roll}) to the next--to--leading term ${\cal O}(x^{3/2}/z)$:

\begin{eqnarray}
&&\epsilon=1+4\frac{x}{z}+2 {\rm sign} (\alpha z)\sqrt{\frac{6}{\kappa}} |\gamma| \frac{x^{3/2}}{z}+{\cal O}(\frac{x}{z^2})\nonumber\\
&& \Rightarrow (\epsilon-1)z\rightarrow 4x+2 {\rm sign}(\alpha z)\sqrt{\frac{6}{\kappa}}|\gamma|x^{3/2},
\end{eqnarray}
in order to get $x^{\prime}$ from~(\ref{eq:x_prime}). Moreover, setting $t=1/z$ in Eqs.~(\ref{eq:x_prime}, \ref{eq:z_prime}) 
the two variables $(x,t)$ decouple close to the critical point $(x,t)=(0,0)$


\begin{eqnarray}
    x'&=& 2 {\rm sign}(\alpha t)\sqrt{\frac{6}{\kappa}}|\gamma|x^{3/2}\label{eq:x_prime_infty},\\
    t'&=& 2 t.
    \label{eq:t_prime_infty}
\end{eqnarray}

\noindent The equation above shows that $(x,t)=(0,0^{-})$ ($(x,z)=(0,-\infty)$) is an attractor when ${\rm sign}(\alpha)<0$ and a saddle point when ${\rm sign}(\alpha)>0$. In Fig.~\ref{fig:critical_points} this is schematically represented by the arrows pointing either toward the negative $z$ axis or away from it. A summary of the critical points discussed in the last two Sections is provided in Table~\ref{table:critical_points}.

\begin{table}[t]
\begin{center}
{\begin{tabular}{|@{}|c|c|c|c|@{}}
\hline
   &   $(x,z)$ & $0<|\gamma|<\sqrt{6\kappa}$ & $\sqrt{6\kappa}<|\gamma|<\infty$\\
\hline
Kination & $(1,0)$ & Saddle point &  \begin{tabular}[c]{c}Attractor ($\alpha z<0$) $\circled{3}$, $\circled{7}$, $\circled{8}$\\ Saddle point ($\alpha z>0$)\end{tabular} \\
\hline
GB & $(0,1)$ & \multicolumn{2}{c|}{Saddle point} \\
\hline
Fast roll & $(x_{\rm fp},z_{\rm fp})$ & Attractor ($\alpha>0$)$\circled{5}$,$\circled{6}$ &  \begin{tabular}[t]{c}Saddle point ($\alpha<0$,$|\gamma|<\sqrt{8\kappa}$)\\ Repulsive node ($\alpha<0$,$|\gamma|>\sqrt{8\kappa}$)\end{tabular}\\
\hline
Slow roll & $(0,-\infty)$ & \multicolumn{2}{c|}{\begin{tabular}[t]{c}Attractor ($\alpha<0$)$\circled{1}$, $\circled{2}$,$\circled{4}$\\ Saddle point ($\alpha>0$)\end{tabular}} \\
\hline
\end{tabular}}
\caption{Summary of the critical points discussed in Sections~\ref{sec:finite_critical_points} and~\ref{sec:attractors_at_infinity}.
Next to each attractor the circled number indicates the flows shown schematically in Fig.~\ref{fig:critical_points}. The same numbers are also shown in the different domains of Fig.~\ref{fig:eos_alpha_gamma} for the equation of state.
  \label{table:critical_points}}
\end{center}
\end{table}

\subsection{Flows of the autonomous equation}
\label{sec:flows}

\begin{figure*}[htb!]
\centering
\includegraphics[width=6cm,height=6.2cm]{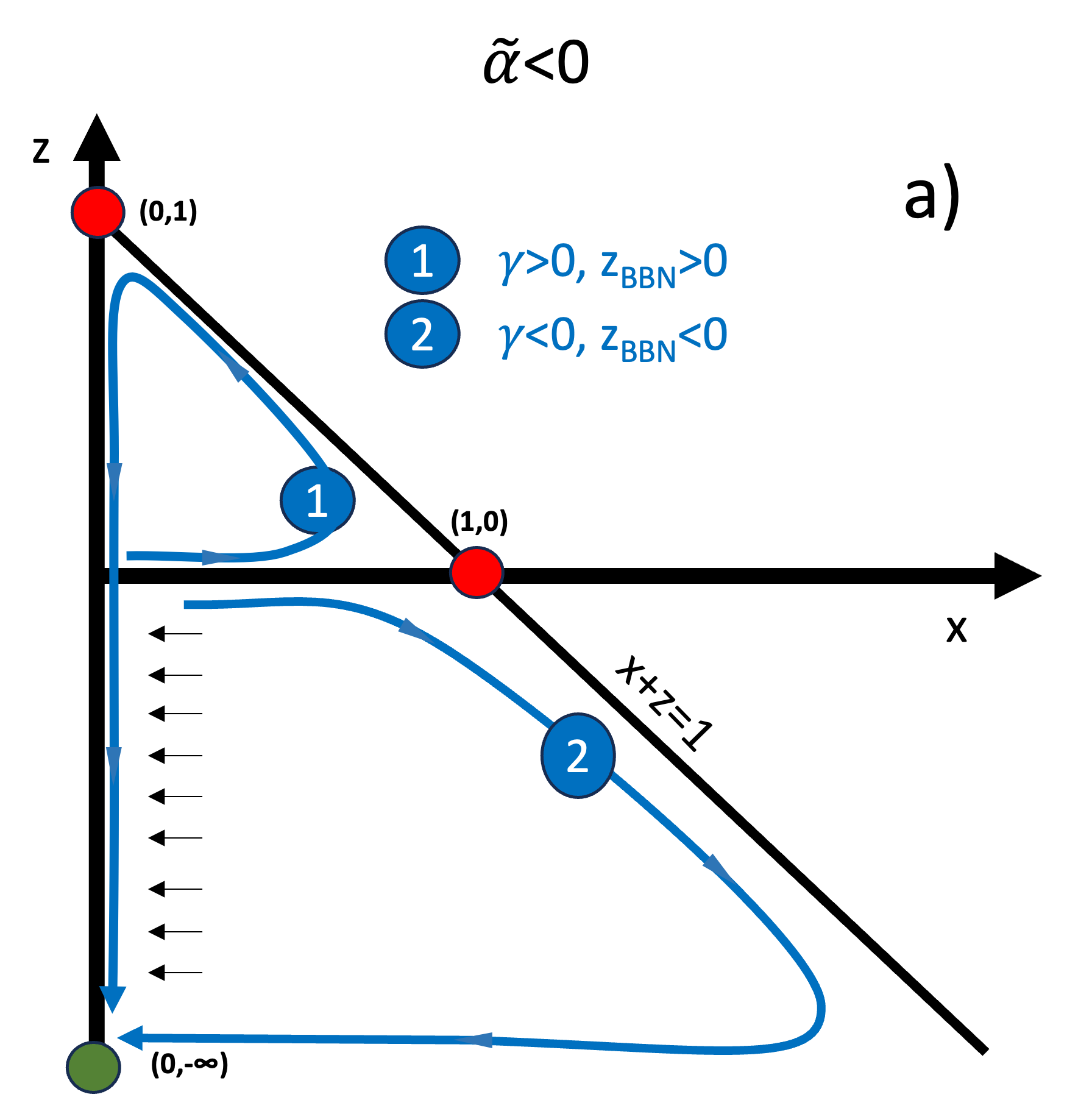}
\includegraphics[width=7.5cm,height=6.2cm]{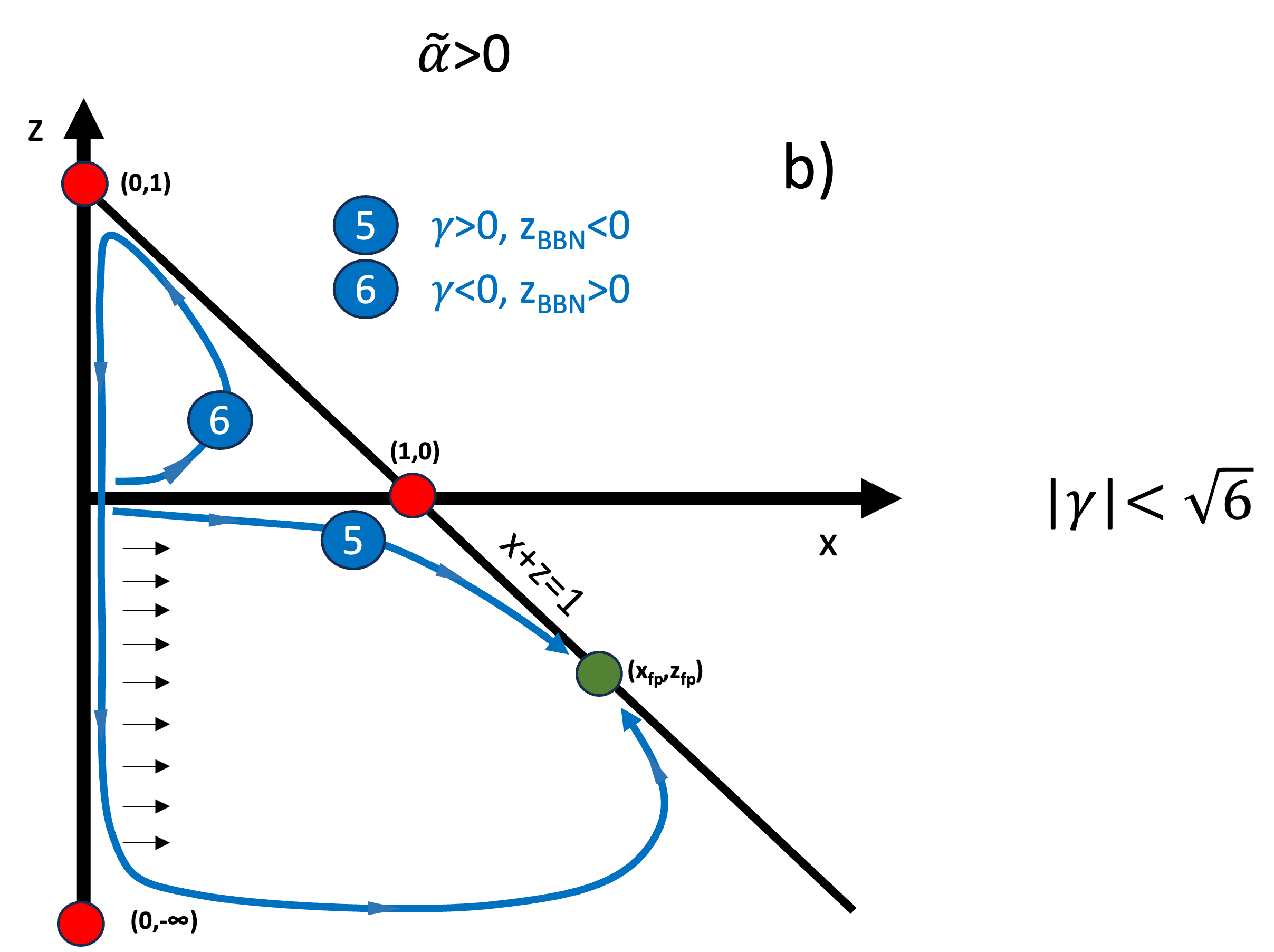}
\includegraphics[width=6cm,height=6.2cm]{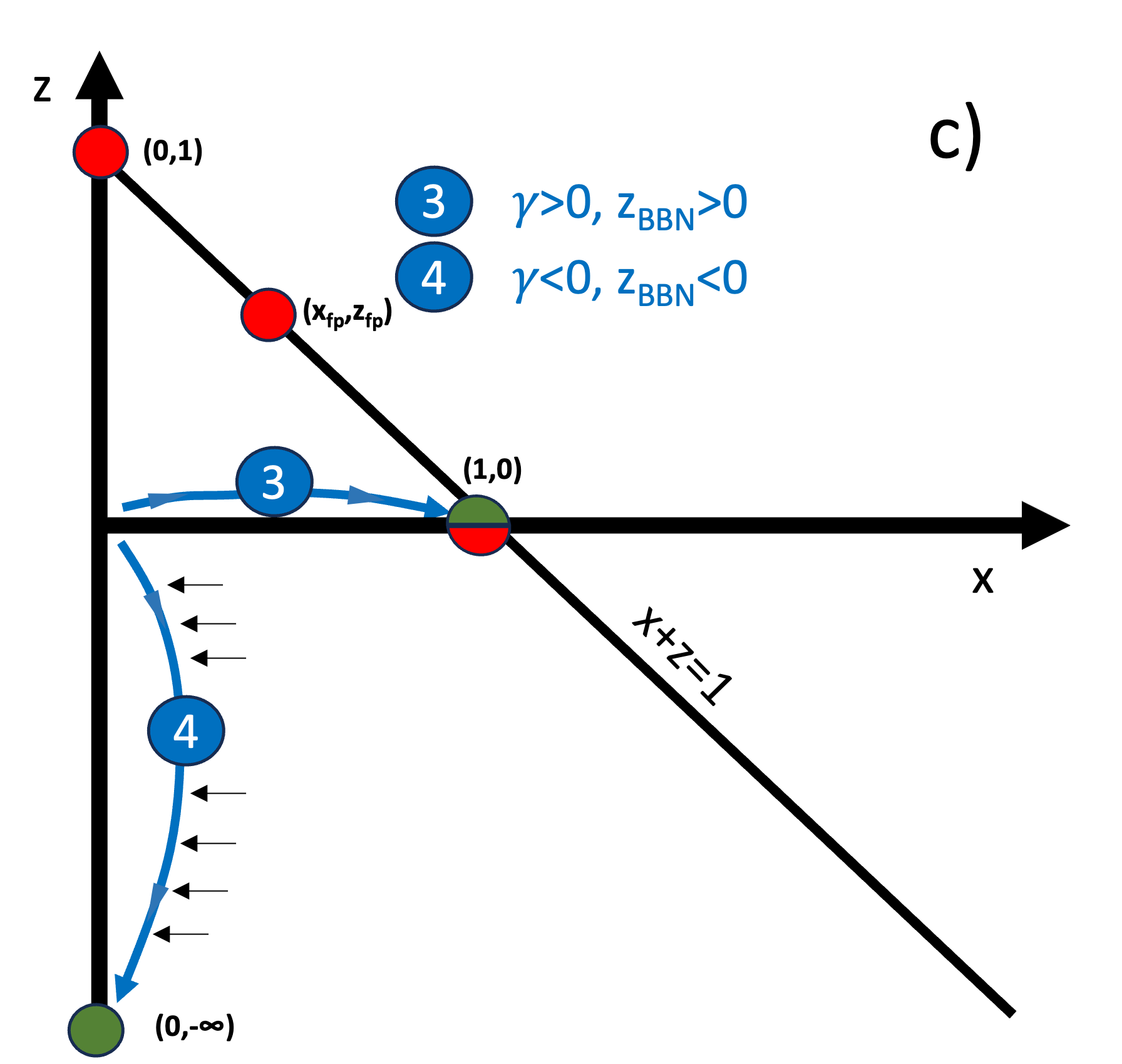}
\includegraphics[width=7.5cm,height=6.2cm]{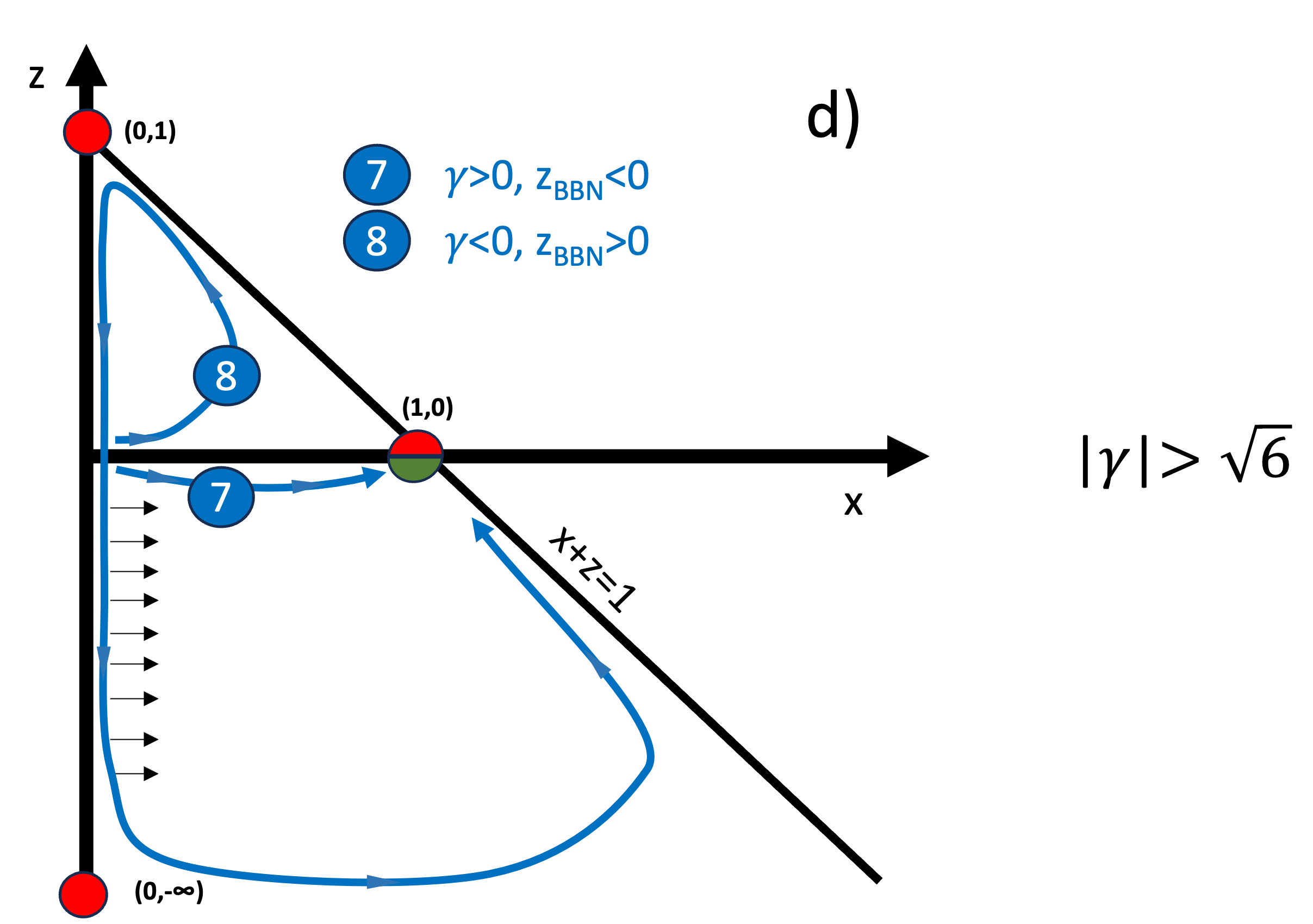}
\includegraphics[width=5cm]{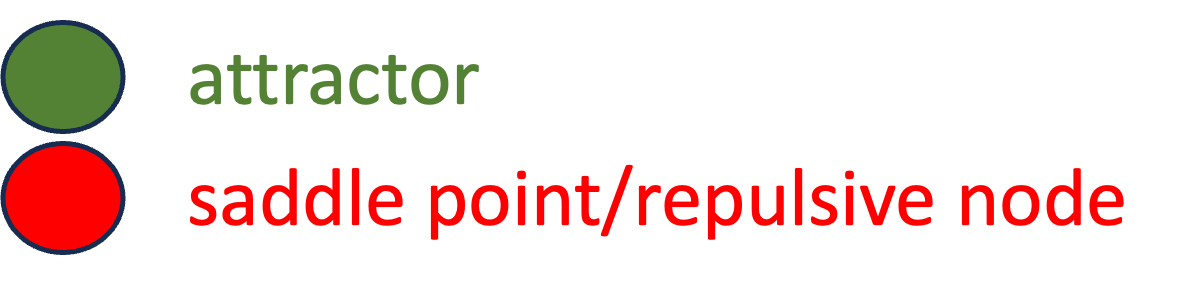}
\caption{Schematic description of the critical points and of the qualitative evolution of the autonomous differential equations~(\ref{eq:x_prime}, \ref{eq:z_prime}) (in the caption $\kappa$ =1). Circles filled in green or red represent stable and unstable critical points, respectively. In the case of $(x_c,z_c)=(1,0)$ only half of the circle is filled in green, depending on whether the point is an attractor for $z\rightarrow 0^+$ or $z\rightarrow 0^-$ (see text). The flows numbered from $\circled{1}$ to $\circled{8}$ correspond to the eight paths that exhaust the possible high--temperature evolution of dEGB Cosmology. The same numbers are shown in Fig.~\ref{fig:eos_alpha_gamma}.}
\label{fig:critical_points}
\end{figure*}

The asymptotic behaviour of the equation of state shown in Fig.~\ref{fig:eos_alpha_gamma} has a clear interpretation in the discussion of Sections~\ref{sec:finite_critical_points} and~\ref{sec:attractors_at_infinity} about the attractors of the autonomous system of differential equations~(\ref{eq:x_prime}, \ref{eq:z_prime}). In particular, the plots of Fig.~\ref{fig:critical_points} provide a qualitative explanation of how the different attractors are approached depending on the boundary conditions. 

We start our discussion by analysing the behaviour of the differential equations in the origin. In particular, one can notice that the mapping between the differential equations of Eqs.~(\ref{Eqfrw1}, \ref{Eqfrw2}, \ref{Eqfrw3}) and those of Eqs.~(\ref{eq:x_prime}, \ref{eq:z_prime}) presents there a degeneracy, since the $\epsilon$ function defined in Eq.~(\ref{eq:epsilon_xz}) is undetermined for $(x,z)$ = $(0,0)$ . In particular, from the point of view of Eqs.~(\ref{eq:x_prime}, \ref{eq:z_prime}) the two variables $x$ and $z$ are independent, and the value of $\epsilon$ in the origin depends on the specific flow line followed by the system to approach it\footnote{In the origin an infinite set of different flows of the autonomous equations~(\ref{eq:x_prime}, \ref{eq:z_prime}) converge. However each solution through the origin is unique for a given set of boundary conditions (i.e. Eqs.~(\ref{eq:x_prime}, \ref{eq:z_prime}) satisfy the Cauchy–Lipschitz theorem~\cite{Coddington_Levinson_1955}).}. However, from their definition~(\ref{eq:def_xyz}) it is clear that $x$ and $z$ are correlated in the origin, since they can only vanish together when $\dot{\phi}\rightarrow$ 0. In particular, to calculate $\epsilon$ in the origin one can use the explicit expressions of $x$ and $z$ in terms of $\rho_{\phi}$ and $\rho_{\rm GB}$:

\begin{eqnarray}
&&\lim_{x,z\rightarrow 0} \epsilon(x,z)=\frac{8+\frac{z^2}{x}}{4+\frac{z^2}{x}},\label{eq:epsilon_limit}\\
&&\frac{z^2}{x}=\frac{\rho_{\rm GB}^2}{\rho\rho_{\phi}}= 384 \kappa \left (f^{\prime}\right )^2 H^4\simeq (8\pi)^2 0.14 g_{*}^2\left (\frac{\alpha}{\mbox{km$^2$}}\right)^2\gamma^2 e^{2\gamma\phi}\left (\frac{T}{\mbox{GeV}}\right )^8,\label{eq:z2_over_x} 
\end{eqnarray}

\noindent which shows that (i) in the origin $\epsilon$ is well defined, since in the $z^2/x$  combination $\dot{\phi}\rightarrow 0$ cancels out (this needs to be the case since nothing special happens for $\dot{\phi}=0$ in the original Friedmann equations (\ref{Eqfrw1}, \ref{Eqfrw2}, \ref{Eqfrw3})); (ii) the actual value of $\epsilon(0,0)$ depends on the previous evolution history of the system; (ii) the quantity $z^2/x$ is extremely sensitive to the temperature.  At this stage it is convenient to divide the discussion in two cases, depending on the value of $\rho_{\phi}(T_{\rm BBN})$. 

\begin{itemize}
    \item{\bf $\rho_{\phi}(T_{\rm BBN}) = 0$.}  

    When $\dot{\phi}$ is set to zero at BBN ($T$ = 1 MeV) the flow originates from the (0, 0) point and in Eq.~(\ref{eq:z2_over_x}) $z^2/x\ll 1$ so that $\epsilon\rightarrow 2$ and $w=1/3$, i.e. standard Cosmology is recovered. Moreover, Eq.~(\ref{eq:z_prime}) yields $(z^{\prime})_{\rm BBN}\simeq z^2/(2x)>0$, and since we evolve the differential equations in the negative e--folding direction this implies that $z$ grows negative. In other words, if the system starts from the origin it can only evolve to the negative $z$ plane, following one of the paths that in Fig.~\ref{fig:critical_points} are indicated with the numbers $\circled{2}$, $\circled{5}$, $\circled{4}$ and $\circled{7}$ (this explains why in the last row of Fig.~\ref{fig:eos_alpha_gamma}, corresponding to  $\rho_{\phi}(T_{\rm BBN})=0$, only such numbers appear). For all combinations of $\tilde{\alpha}$ and  $\gamma$, there is always an attractor in the $z<0$ plane, so the system quickly converges to it. In particular, for $\tilde{\alpha}<0$ the attractor is the slow-roll solution $(0,-\infty)$ with $w=-1/3$, while for $\tilde{\alpha}>0$ the attractor is either $(x_{\rm fp},z_{\rm fp})$ with $1<w<7/3$ when $|\gamma|<\sqrt{6\kappa}$ (fast roll) or $(1,0)$ with $w=1$ when $|\gamma|>\sqrt{6\kappa}$ (kination). As already pointed out, such results are symmetric with respect to the sign of the parameter $\gamma$, as it is observed, for instance, in the plots of the last row of Fig.~\ref{fig:eos_alpha_gamma}.

    \item{\bf $\rho_{\phi}(T_{\rm BBN})>0$.}

    In this case, the flow originates from $(x_{\rm BBN},z_{\rm BBN})$, with the initial sign of $z_{\rm BBN}$ depending on the boundary conditions. In particular for our choice $\dot{\phi}(T_{\rm BBN})>0$ for $\tilde{\alpha}\gamma>0$ one has $z_{\rm BBN}<0$ and qualitatively the evolution of the system is very similar to the case for $\rho_{\phi}(T_{\rm BBN})=0$ already discussed (the boundary conditions at BBN ``go with the flow", see trajectories $\circled{2}$, $\circled{5}$, $\circled{4}$ and $\circled{7}$ in Fig.~\ref{fig:critical_points}). This is confirmed by the plots of Fig.~\ref{fig:eos_alpha_gamma}, that indeed when $\dot{\phi}(T_{\rm BBN})>0$ for $\tilde{\alpha}\gamma>0$ look similar to those for which $\dot{\phi}(T_{\rm BBN})=0$. On the other hand, a different situation emerges when $\tilde{\alpha}\gamma<0$ and $z_{\rm BBN}>0$. From Table~\ref{table:critical_points} and Fig.~\ref{fig:critical_points} one can see that in most cases the only attractor of the system is in the negative $z$ plane. In such cases the boundary conditions at BBN ``kick the flow in the wrong direction'' so that after a transient path in the $z>0$ plane (see paths $\circled{1}$, $\circled{6}$ and $\circled{8}$ in Fig.~\ref{fig:critical_points}) the system needs to evolve from the positive $z$ plane to the negative $z$ plane. This can only happen by crossing the origin and explains the cases where $\dot{\phi}$ is observed to change sign at high temperature (in  Figs.~\ref{fig:rhos_gamma1} and~\ref{fig:rhos_gamma5} this happens when both $\rho_{\phi}$ and $\rho_{\rm GB}$ vanish). In such situations the system is required to take a large ``detour" in the positive plane before reaching a trajectory tangent to the $z$ axis that allows it to cross the origin. According to Eq.~(\ref{eq:z2_over_x}) now when $\dot{\phi}\rightarrow 0$ one has $z^2/x\gg 1$ and $\epsilon\rightarrow 1$, i.e. the equation of state corresponds to slow-roll, $w=-1/3$. For $\tilde{\alpha}<0$ and $\gamma<\sqrt{6\kappa}$ this is also the only available attractor, so $w$ does not change anymore. On the other hand, for $\tilde{\alpha}>0$ slow-roll is a saddle point, so that $w=-1/3$ is only a metastable solution: the system follows  $w=-1/3$ for a while and then eventually jumps to the real attractor at higher temperatures, which is $(x_{\rm fp}, z_{\rm fp})$ for $|\gamma|<\sqrt{6\kappa}$ and $(1,0)$ for $|\gamma|>\sqrt{6\kappa}$\footnote{Indeed, in the analysis Ref.~\cite{GB_WIMPS_sogang} the evolution of the  dEGB Friedmann equations was limited to the temperatures relevant for WIMP thermal decoupling, and the case $\tilde{\alpha}>0$, $\gamma<0$ was mistakenly assumed to have the slow-roll asymptotic behaviour, missing the transition to a different $w$ at higher temperatures.}. As far as the asymptotic equation of state is concerned, this implies that for $\tilde{\alpha}>0$ the plots of Fig.~\ref{fig:eos_alpha_gamma} are very similar to the $\dot{\phi}(T_{\rm BBN})=0$ case. The only exception to this behaviour corresponds to $\tilde{\alpha}<0$ and $\gamma>\sqrt{6\kappa}$. In this case, a new attractor appears at $(1,z\rightarrow 0^+)$, i.e. in the {\it positive} plane, so that, instead of crossing the origin and reaching slow--roll, the system quickly asymptotizes to kination (see track $\circled{3}$ in Fig.~\ref{fig:critical_points}). This explains why in Fig.~\ref{fig:eos_alpha_gamma} for $\tilde{\alpha}<0$ the equation of state turns from slow-roll ($w=-1/3$) to kination ($w=1$) for $\gamma>\sqrt{6\kappa}$ when $\dot{\phi}(T_{\rm BBN})>0$, while it is always slow-roll when $\dot{\phi}(T_{\rm BBN})=0$. We conclude by pointing out that also kination can be a metastable regime, for those trajectories that approach the $(1,0)$ saddle point before eventually moving to another attractor (see tracks $\circled{1}$, $\circled{2}$, $\circled{5}$, $\circled{6}$ and $\circled{8}$ in Fig.~\ref{fig:critical_points}). In such case  the system follows  $w=1$ for a while and then eventually jumps to the real attractor at higher temperatures.
    The appearance of such meta--stable solutions for the equation of state, which are clearly visible in Fig.~\ref{fig:eos} and also in Fig.~\ref{fig:eos_neg_gammas}, 
    can have important consequences for the GW stochastic background discussed in Section~\ref{sec:GW_constraints}, which depends not only on the asymptotic behaviour at high temperature but on the whole expansion history of the Universe. 
\end{itemize}

\section{Constraints from Gravitational Waves}
\label{sec:GW_constraints}

\subsection{GW stochastic background}
\label{sec:GW_stochastic}

\subsubsection{Theoretical framework}
\label{sec:GW_theory}
Any plasma of relativistic particles in thermal equilibrium emits a stochastic background of gravitational waves, which in the case of the Standard Model is expected to peak at a frequency of around 80 GHz~\cite{Ghiglieri:2015nfa,Ghiglieri:2020mhm}. At variance with other hypothetical GW sources such as scenarios of inflation, reheating models and phase transitions in GUT theories, such process is an irreducible source of GWs that only involves standard physics, and potentially represents an unambiguous way to directly probe Cosmological models before Big Bang Nucleosynthesis.  So, although present detectors are only sensitive to frequencies of the order of few Hertz, some proposals exist to extend the experimental reach to the GHz range~\cite{Ito:2019wcb, Herman:2022fau}. 

The magnitude and spectral shape of the corresponding stochastic GW background that is produced during the history of the Universe from the beginning of the thermal--radiation dominated
epoch after the big bang until the electroweak crossover, at a temperature $T_{\rm EWCO}$ = 160 GeV, has been calculated in~\cite{Ghiglieri:2015nfa, Ghiglieri:2020mhm}. At a given time it can be expressed as
\bea
\frac{1}{a^4}\frac{d}{dt}\left (a^4 \rho_{\rm GW}(t) \right )=\left (\frac{\partial}{\partial t}+ 4 H \right)\rho_{\GW}(t)=4 \frac{T^4}{\MpR^2} \, \int \frac{d^3k}{(2\pi)^3}   \eta(T,k),
\label{eq:evolvgw2}
\eea
where $\eta(k,T)$ is the shear viscosity of the SM plasma. The equation above neglects the back reaction contribution from gravitational excitations annihilating or decaying back into the plasma, which is a good approximation for temperatures below the Planck scale. In Eq.~(\ref{eq:evolvgw2}) the right--hand side  encodes the information related to the GW production, which is modeled by the function $\eta(k,T)$, while the left--hand side describes the evolution of $\rho(t)_{\GW}$ according to the Cosmological model and, in particular, to the temperature dependence of the Hubble parameter. 

For a plasma kept in thermal equilibrium by SM interactions, $\eta(k,T)$ is a function of $\hat{k}=k/T$ and $T$ given by
\bea
\label{eq:eta_complete_conts}
\eta(\hat{k},T)  
&=&
 \left\{
\begin{array}{cc}
\frac{1}{8\pi}\frac{16}{g_1^4 \ln(5T/m_{D_1})}, & \, k \lesssim  \alpha_1^2 T,\\
\eta_{\rm{HTL}}(\hat{k},T)+ \eta^T(\hat{k},T),
& k  \gtrsim 3 T,
\end{array}
\right.,
\eea
\noindent where $\alpha_1=g_1^2/4\pi$ with $g_1$ the weak--hypercharge coupling and $\eta_{\rm{HTL}}(\hat{k},T)$ is the so called Hard Thermal Logarithmic (HTL) expression~\cite{Braaten:1989mz,Ghiglieri:2020mhm}:
\bea
\label{eq:etaHTL}
\eta_{\rm{HTL}}(\hat{k},T)=\frac{1}{16 \pi} \hat{k} n_B(\hat{k})\, \sum^3_{i=1} \hat{m}^2_{D_i} \ln\left(4\frac{1}{\hat{m}^2_{D_i}} {\hat{k}}^2+ 
1 \right),
\eea
where $n_B(\hat{k})=\frac{1}{e^{\hat{k}}-1}$, and the Debye masses are given by
\bea
\hat{m}_{D_i}=\frac{m_{D_i}}{T},\quad
m^2_{D_i}=\left\{ 
\begin{array}{c}
d_1 \frac{11}{6} g_1^2 T^2, \, d_1=1, \\
\\
d_2 \frac{11}{6} g_2^2 T^2, \,
d_2=3\!\! \\
\\
d_3 2 g_3^2 T^2, \,
d_3=8. \\
\end{array}
\right.,
\eea
In Eq.~(\ref{eq:eta_complete_conts}) $\eta^T(\hat{k},T)$ represents the contribution from the thermal corrections computed in \cite{Ghiglieri:2020mhm} and is given by
\bea
\label{eq:etaThermal}
\eta^T(\hat{k},T)&=& \left[3 g_2^2(T)+ 12 g_3^2(T)\right]\eta_{gg}(\hat{k}) +
\left[g_1^2(T) + 3 g_2^2(T) \right]\eta_{sg}(\hat{k}) \nonumber\\
&&+ \left[5g_1^2(T)+ 9 g_2^2(T) + 24 g_3^2(T)\right]\eta_{fg}(\hat{k}) \nonumber\\
&&+ \left[3 |y_t(T)|^2 + 3 |y_b(T)|^2+ |y_\tau(T)|^2\right] \eta_{sf}(\hat{k}).
\eea
\noindent
In the expression above, the couplings $g_i$ of the SM gauge group $SU(3)_{\rm{c}} \times SU(2)_{\rm{L}}\times U(1)_{\rm{Y}
} $  and the Yukawa couplings $y_t$, $y_b$ and $y_\tau$ depend on the temperature, with $y_b$ and $y_\tau$ giving a negligible contribution. In Appendix~\ref{app:thermalcorr}, we provide the functions $\eta_{gg}(\hat{k})$, $\eta_{sg}(\hat{k})$, $\eta_{fg}(\hat{k})$ and $\eta_{sf}(\hat{k})$ in the notation of \cite{Ringwald:2020ist}, and correct a typo in one of them.

In the next Section, we will use the numerical solutions discussed in Section~\ref{sec:numerical} for the Friedmann equations of the dEGB cosmological model to calculate numerically the following quantity~(see for example \cite{Maggiore:2018sht}):
\begin{equation}
    \Omega(f, T_0)h^2\equiv\frac{1}{\rho_{\crit}(T_0)}\frac{d \rho_{\rm GW}(T_0)}{d\ln f} h^2,
    \label{eq:omega_f}
\end{equation}
\noindent where $h=H_0/(100\,\mbox{km/s/Mpc})$ with $H_0$ the present value of the Hubble constant, $T_0=2.7\,K$ is the present CMB temperature, $f$ the frequency measured today and $\rho_{\crit}=3H^2/(8\pi G)$ the critical density corresponding to a flat Universe. Note that the quantity $\Omega(f, T_0)h^2$ is independent of the actual Hubble expansion rate.

In order to calculate Eq.~(\ref{eq:omega_f}), we follow Refs.~\cite{Ghiglieri:2015nfa, Ghiglieri:2020mhm} and proceed in two steps. First, we integrate Eq.~(\ref{eq:evolvgw2}) from some temperature $T_{\rm max}$ (that we will identify with the reheating temperature $T_{\rm RH}$ at the end of inflation) and $T_{\rm EWCO}$ =160 GeV to obtain $d\rho_{\rm GW}/d\ln k_{\rm EWCO}$, where $k_{\rm EWCO}= k(T)\left (g_{*s}(T_{\rm EWCO})/g_{*s}(T) \right )^{1/3} T_{\rm EWCO}/T$ is the wave number at $T$ = $T_{\rm EWCO}$. In order to do this, in Eq.~(\ref{eq:evolvgw2}) we change variable from $t$ to $T$ by making use of Eq.~(\ref{eq:T-t}). In a second step, we rescale the GW density from $T=T_{\rm EWCO}$ to $T$ = $T_0$ with the scale factor ratio  $(a_{\rm EWCO}/a_0)^4$ and we use
\begin{equation}
    f=\frac{1}{2\pi}\left [ \frac{g_{\rm *s}(T_0)}{g_{\rm *s}(T_{\rm EWCO})}\right]^{\frac{1}{3}}\left (\frac{T_0}{T_{\rm EWCO}} \right ) k_{\rm EWCO},
    \label{eq:f_k_ewco}
\end{equation}
\noindent to express the frequency $f$ in terms of $k_{\rm EWCO}$, and normalize the GW spectrum to the present photon density $\Omega_{\gamma,0}=2.4729(21)\times 10^{-5}$ writing $h^2/\rho_{\crit}(T_0)=15/(\pi^2 T_0^4)\Omega_{\gamma,0}$. The final result is

 \bea
 \label{eq:FinalformOmegaGW}
 {\Omega_{\rm {GW}}(f,T_0) h^2 }&=&{\Omega_{\gamma_0} h^2}
\frac{\lambda}{\Mp}\int^{T_{\rm Max}}_{T_{\rm EWCO}}\, dT \left(\frac{g_{*0}}{g*(T)} \right)^{4/3} \, T^2\, \hat{k}(f,T)^3 \frac{\eta(\hat{k},T)}{\sqrt{\rho(T)}}\, \beta(T),
 \eea
where $\Omega_{\rm {GW}}(f,T_0)$ is the fraction of energy liberated into  gravitational wave radiation per frequency octave~\cite{Kamionkowski:1993fg, Huber:2008hg, Lee:2021nwg}, $\lambda=30\sqrt{3}/\pi^4$, $\beta(T)$ is given by Eq.~(\ref{eq:beta_T_t}), $g_{*0}$ = 2, $g_{*s}(T_0)$ = 3.91, $g_{*s}(T_{\rm EWCO})$ = 106.75, and

\begin{equation}
    \hat{k}(f,T)=\left [\frac{g_{*s}(T)}{g_{*s}(T_0)} \right ]^{\frac{1}{3}}\frac{2\pi f}{T_0   }.
\end{equation}

At production, the GW source term $\eta(\hat{k},T)$ has a peak at $\hat{k}\simeq$ 3.92~\cite{Ghiglieri:2020mhm,Ringwald:2020ist} 
that is largely independent of the temperature, 
and whose position in the frequency $f$ measured today corresponds to $f_{\rm peak}\simeq74\, \mbox{GHz}$, assuming the particle content of the Standard Model for $T>T_{\rm EWCO}$.

\subsubsection{Present and future stochastic GW observations}
\label{sec:GW_experiment}
Presently, the only bound on the GW stochastic background discussed in the previous Section can be obtained from BBN, $\Omega(f, T_0)h^2<1.3\times 10^{-6}$~\cite{Yeh:2022heq}. In Figs.~ \ref{fig:GWconstEx1}-\ref{fig:GW_alpha_positive_gamma_negative} this bound is represented by the horizontal solid line. We have also plotted a forecasted \cite{Pagano:2015hma} future sensitivity (dotted line), $O(10^{-7})$, achievable with future satellite missions such as COrE \cite{thecorecollaboration2011core} and EUCLID \cite{laureijs2011euclid}.
On the other hand, the current and planned sensitivity of GW experiments is far below the GHz frequency range, although some experimental ideas have the potential to reach it. For instance, an interesting possibility is the one proposed in~\cite{Ito:2019wcb} where gravitational waves can be observed by using the resonant quantum excitations of spin waves (magnons) produced inside a microwave cavity. However, the size of the cavity, $\simeq 5$ cm, is about one order of magnitude bigger than the one required to match the GW peak wavelength, and also its current sensitivity is not competitive with the BBN limit. Another interesting proposal is that of \cite{Herman:2022fau}, where it has been suggested that GWs  can be detected using resonant electromagnetic cavities. Currently, such detectors are only sensitive to frequencies below about 100 MHz. However, in~\cite{Herman:2022fau} the projected sensitivity to the characteristic strain $h^2_c(f)$, which has an oscillatory behaviour in the frequency interval between $3.5\times 10^7$ Hz and $10^9$ Hz, seems to reach a plateau at higher frequencies that can potentially constrain the lower end of the stochastic GW background from the SM plasma. This can be seen in~ Figs.~\ref{fig:GWconstEx1}--\ref{fig:GW_alpha_positive_gamma_negative}, where we have plotted the projected bound from~\cite{Herman:2022fau} in the density vs frequency plane, showing  the first resonant peak at around $3.5 \times 10^7$ Hz (solid line), and extrapolating the plateau up to 100 GHz (dashed line). In  Figs.~\ref{fig:GWconstEx1}--\ref{fig:GW_alpha_positive_gamma_negative} we have converted the bound on $h^2_c(f)$ from Fig.~1 of \cite{Herman:2022fau} using $\Omega_{\rm {GW}}(f)=(2\pi^2/3H_0^2)f^2 h^2_c(f)$ (see for example \cite{Kuroda:2015owv} ).  Indeed, below $10^9$ Hz such constraint appears competitive to the BBN bound and potentially able to constrain the GB scenario. Of course, in order to obtain more robust bounds an extension of the experimental sensitivity to $\sim 100$ GHz will be required.

\subsection{GW constraints from compact binary mergers}
\label{sec:GW_binaries}

In this Section we summarize the bounds from the detection of Gravitational Waves from compact binary mergers and refer to Ref.~\cite{GB_WIMPS_sogang} for further details. 

Due to the Universe expansion even if $\dot{\phi}(T_{\rm BBN})\ne$ 0  the evolution of the scalar field $\phi$ eventually freezes at some asymptotic temperature $T_L\ll T_{\rm BBN}$ to a constant background value $\phi(T_L)$, implying no departure from GR at the cosmological level for $T<T_L$. On the other hand, in the vicinity of a BH or a NS the density profile of the scalar field is distorted compared to $\phi(T_L)$, leading to a local departure from GR that can modify the GW signal if the stellar object is involved in a merger event. Near the black hole or the neutron star the distortion of the scalar field is small and the dEGB function $f(\phi)$ can be expanded up to the linear term in the small perturbation $\Delta \phi$ around the asymptotic value $\phi(T_L)$ of the scalar field at large distance~\cite{BH-NS_GB_2022}  
\begin{equation}
f(\phi) = f\left(\phi(T_L)\right) + f'\left(\phi(T_L)\right) \Delta \phi + {\cal O}\left( (\Delta \phi)^2 \right).
\end{equation}
\noindent In this way, the constraints from compact binary mergers is expressed in terms of $f'\left(\phi(T_L)\right)$:

\begin{equation}
\left| f'\left(\phi(T_L)\right) \right|
\leq \sqrt{8 \pi} \hspace{0.7mm} \alpha^{\rm max}_{\rm GB} , 
\label{eq:GW_constraints}
\end{equation}

\noindent where we take $\alpha^{\rm max}_{\rm GB} = (1.18)^2$ $\rm km^2$ from the analysis of the LIGO-Virgo Collaboration data of Ref.~\cite{BH-NS_GB_2022}. The extra factor of $\sqrt{8 \pi}$ in the equation above
is due to the conversion from the units of Ref.~\cite{BH-NS_GB_2022} (that use $G = c = 1$)  to those of the present work (where $\kappa = 8 \pi G = c = 1$). Taking into account the residual evolution of $\phi$ from $T_{\rm BBN}$ to $T_L$ when $\dot{\phi}(T_{\rm BBN})\ne 0$~\cite{GB_WIMPS_sogang} one obtains the final constraint:

\begin{equation}
\left| \tilde{\alpha} \gamma e^{\gamma \frac{\dot{\phi}_{\rm BBN}}{H_{\rm BBN}}} \right| 
\leq \sqrt{8 \pi} \hspace{0.7mm} \alpha^{\rm max}_{\rm GB}. 
\label{eq:GW_constraints_2}
\end{equation}

\subsection{Summary of GW bounds}
\label{sec:results}

Eq.~(\ref{eq:FinalformOmegaGW}) implies that the gravitational waves produced per e--fold scale as $d\Omega_{\rm GW}/d\ln a\propto T^3/\sqrt{\rho}$. At the same time, since they are produced by the relativistic plasma, a sizeable density of GWs can only be produced when radiation is the dominant component of the energy density, i.e. when $\rho\propto T^4$. As a consequence, we conclude that, necessarily, $d\Omega_{\rm GW}/d\ln a\propto T$, and that it is ultra--violet dominated, i.e. the integral of Eq~(\ref{eq:FinalformOmegaGW}) is driven by the contribution of the GWs emitted at high temperature~\cite{Muia:2023wru}. In the case of a standard Cosmology, in which radiation is assumed to dominate the Universe expansion at all temperatures $T>T_{\rm EWCO}$, this implies that the GW stochastic background is a monotonically growing function of $T_{\rm max}$. This could potentially put bounds on the reheating temperature of the Universe $T_{\rm RH}$, although for standard Cosmology the ensuing stochastic background turns out to be below the BBN bound even for values as high as $T_{\rm RH}\simeq 10^{16}$ GeV~\cite{Muia:2023wru}.  On the other hand, for a non--standard cosmology where radiation dominance stops above some temperature $T_{\rm rad, max}$ the stochastic background is dominated by the gravitational waves produced at $T_{\rm rad, max}$, and increasing $T_{\rm max}$ beyond $T_{\rm rad, max}$ does not change the final result, so that the prospects of detection are even worse than in standard Cosmology. 

The dEGB scenario presents an interesting twist to the picture summarized above. In fact, as discussed in  Sections~\ref{sec:numerical} and~\ref{sec:tracking}, in sizeable parts of its parameter space one of the asymptotic solutions at high temperature corresponds to a ``slow--roll regime" with $w=-1/3$ (see Section~\ref{sec:attractors_at_infinity}) where the energy density of the Universe is dominated by $\rho_{\rm rad},|\rho_{\rm GB}|\propto T^4$ while at the same time $\rho_{\rm rad}+\rho_{\rm GB}\propto T^2$, with a large cancellation between $\rho_{\rm rad}$ and $\rho_{\rm GB}$. As a consequence, this scenario has $d\Omega_{\rm GW}/d\ln a\propto T^2$ and represents an exception to the argument summarized above according to which, necessarily, at most $d\Omega_{\rm GW}/d\ln a\propto T$ during an epoch of sizeable GW production. This is due to the peculiar nature of dEGB, which allows to have and epoch when the relativistic plasma dominates the energy of the Universe while at the same time the rate of dilution with $T$ of the total energy density is slower than what usually expected during radiation dominance\footnote{$\rho_{GB}$ is not a physical component of the energy density of the Universe, so the slow--roll regime is radiation--dominated in the sense that the plasma of relativistic particles is the only physical component of the energy--momentum tensor.}. As we will see this strongly enhances the GW expected signal compared to the standard case and allows to put sensible bounds on the reheating temperature $T_{\rm RH}\lesssim 10^8-10^9$ GeV $\ll 10^{16}$ GeV  in the regions of the dEGB parameter space corresponding to the ``slow roll" asymptotic behaviour.

In light of the discussion above, we can expect an enhancement of the expected GW stochastic background in the regions of the parameter space where the ``slow roll" asymptotic behaviour is achieved, i.e. in those regions that correspond to the flows  $\circled{1}$, $\circled{2}$ and $\circled{4}$ as indicated in Figs.~\ref{fig:critical_points} and~\ref{fig:eos_alpha_gamma}. 

\begin{figure*}[htb!]
\centering
\includegraphics[width=7.49cm,height=6.2cm]{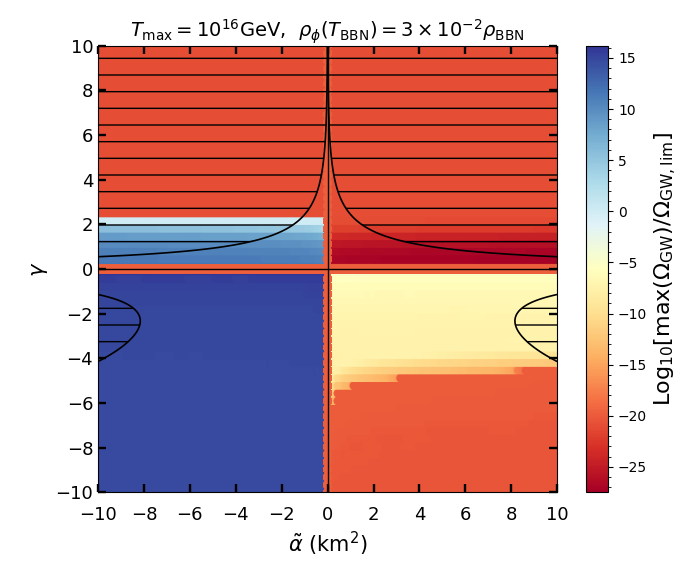}
\includegraphics[width=7.49cm,height=6.2cm]{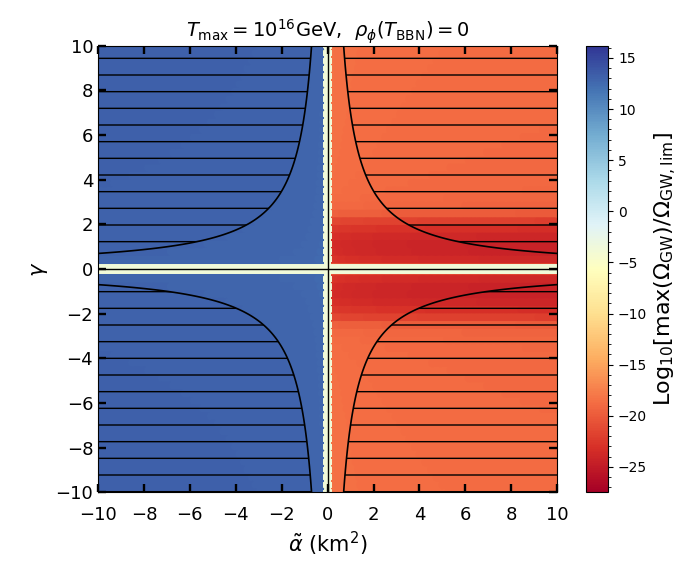}
\includegraphics[width=7.49cm,height=6.2cm]{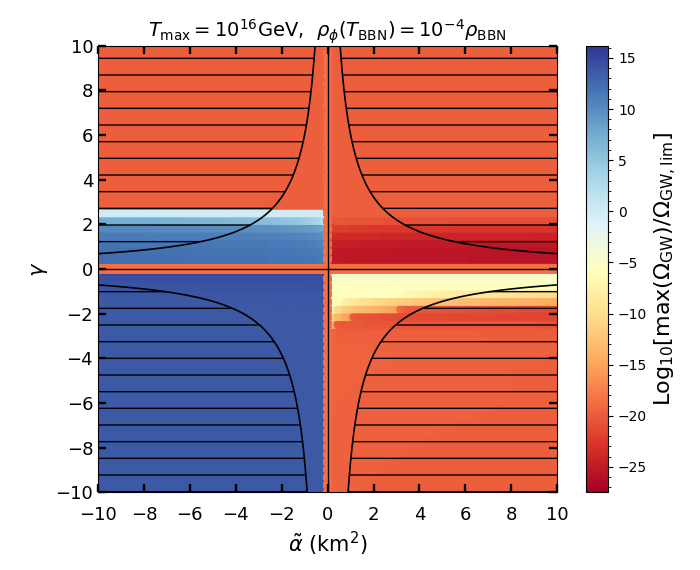}
\caption{The color codes show the variations of the peak value of the GW stochastic background, 
normalised to the BBN upper limit ($\Omega_{\rm GW, lim}h^2 \simeq 10^{-6}$), 
in the $\tilde{\alpha} - \gamma$ plane for different values of 
$\rho_\phi(T_{\rm BBN})$: $3\times10^{-2}\rho_{\rm BBN}$ (top left), $0$ (top right) and $10^{-4}\rho_{\rm BBN}$ (bottom). The maximum temperature of the Universe is assumed to be $10^{16}$ GeV. 
The hatched areas are ruled out by the detection of GW from compact binary mergers.}
\label{fig:GW_signal_alpha_gamma}
\end{figure*}

\begin{figure}
    \centering
    \includegraphics[width=7.3cm]{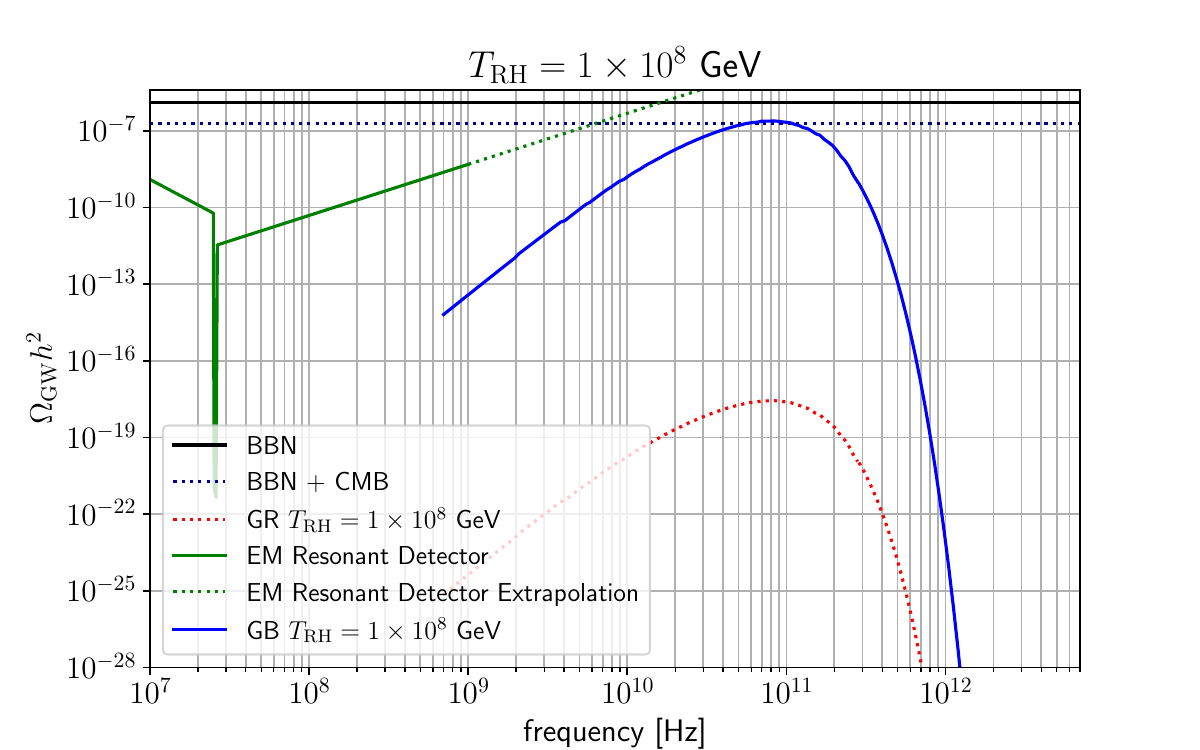}
    \includegraphics[width=7.3cm]
    {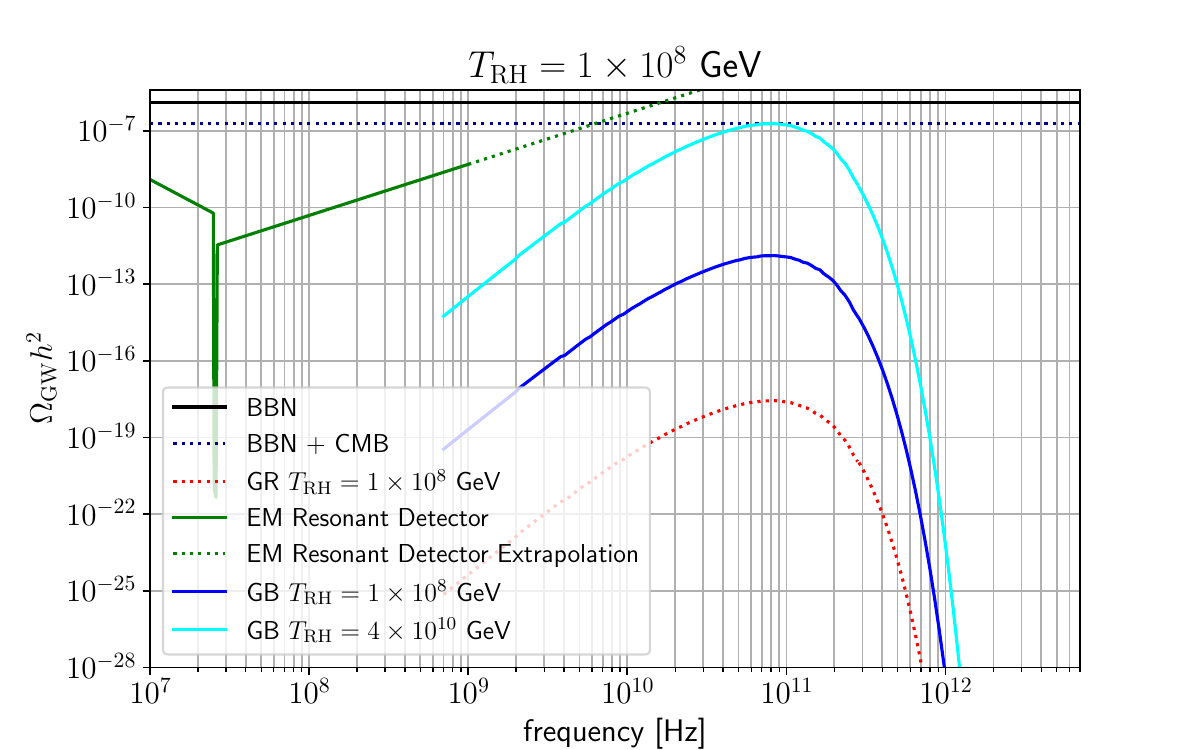}
     \includegraphics[width=7.3cm]
{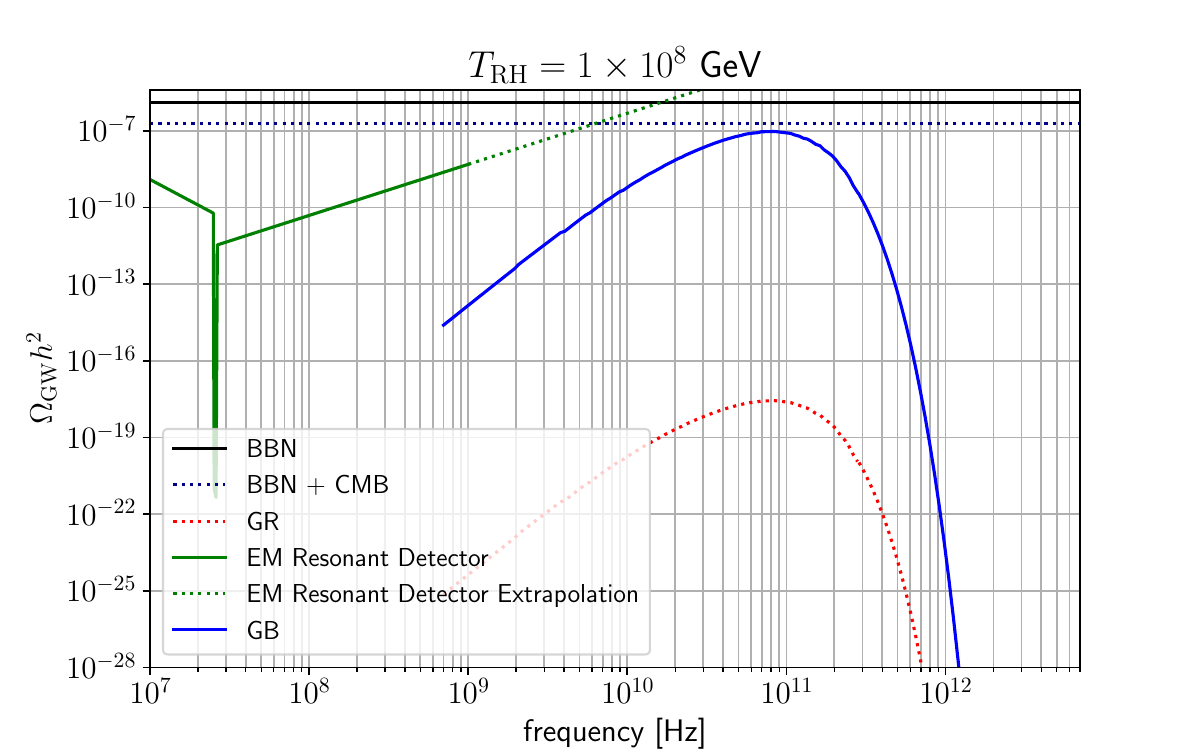}
     \includegraphics[width=7.3cm]
     {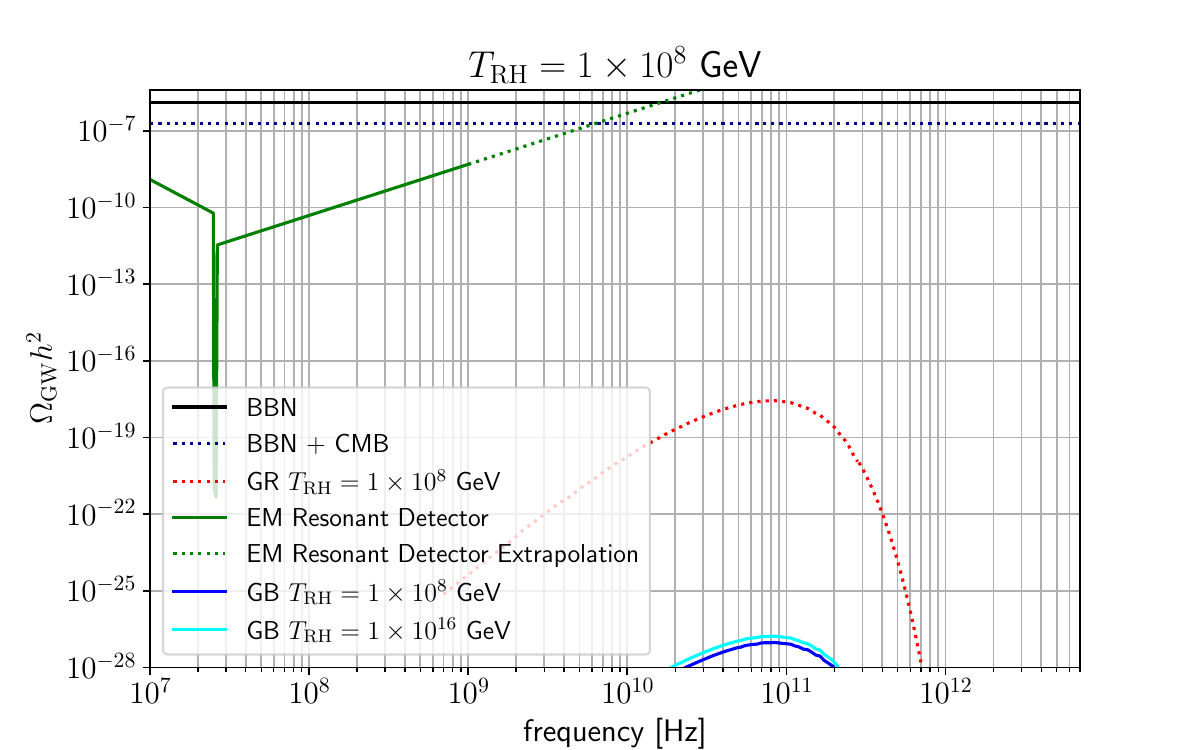}
    \caption{Gravitational Wave density, $\Omega_{\rm{GW}}h^2$, as a function of the frequency, for several dEGB models. The solid line is the current BBN limit~\cite{Yeh:2022heq} while the dotted line is a forecasted sensitivity from the future COrE and EUCLID satellite missions \cite{Pagano:2015hma}. The solid green line corresponds to the resonant detector bound projection of~\cite{Herman:2022fau}, extrapolated to higher frequencies by the dotted line (see Section \ref{sec:GW_experiment}).  In all plots $\tilde{\alpha} = -1 \, {\rm km^2}$. First panel: $\gamma=-1$, second panel:  $\gamma=1$, third panel $\gamma=-5$, fourth panel $\gamma=5$.}
    \label{fig:GWconstEx1}
\end{figure}

\begin{figure}
    \centering    \includegraphics[width=7.3cm]{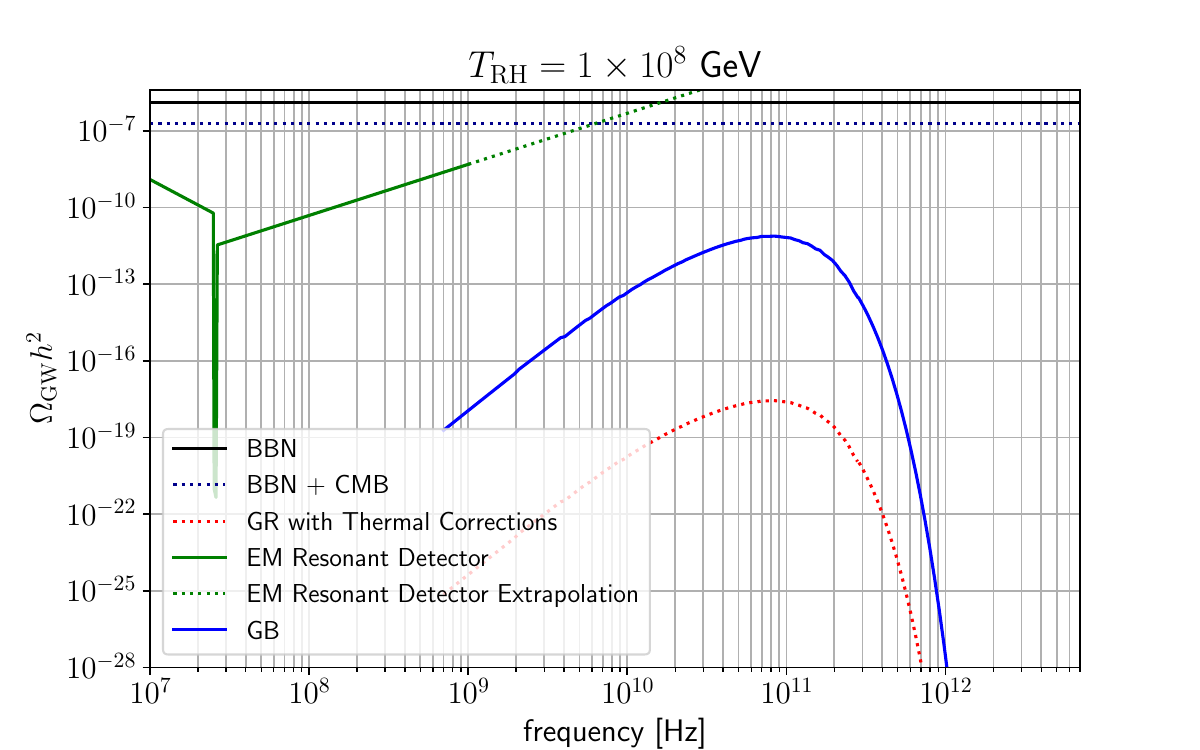}
    \includegraphics[width=7.3cm]{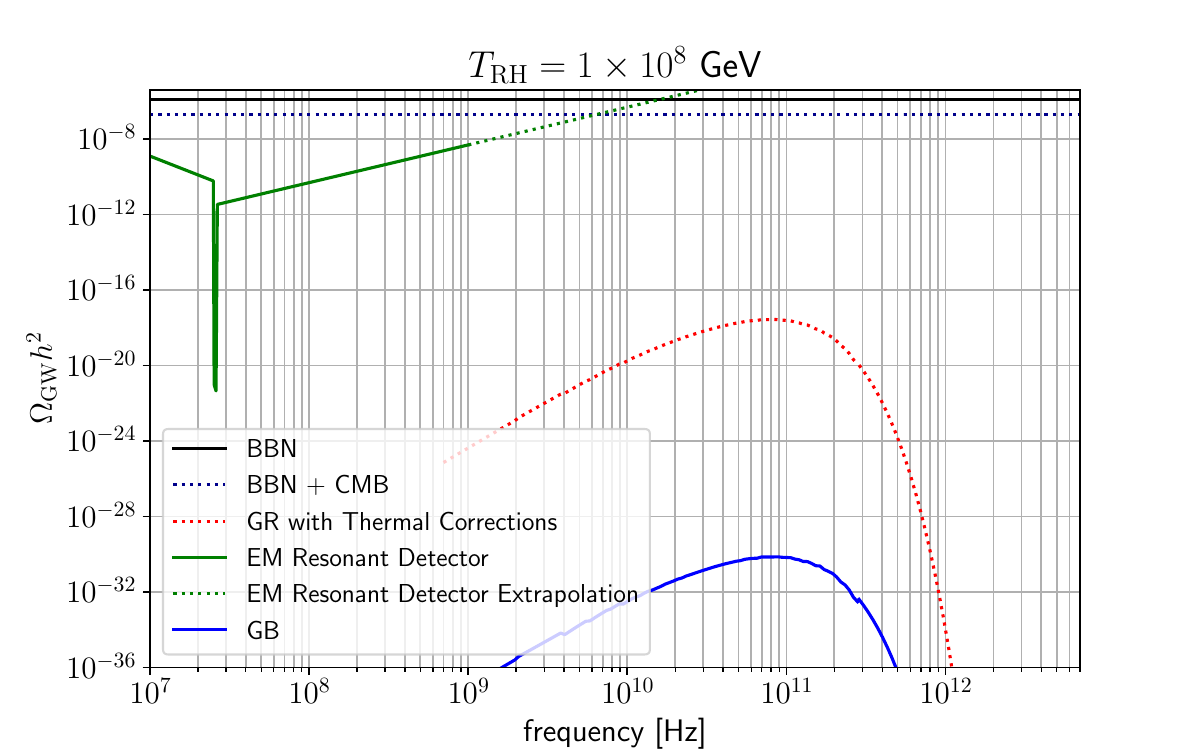}
    \includegraphics[width=7.3cm]{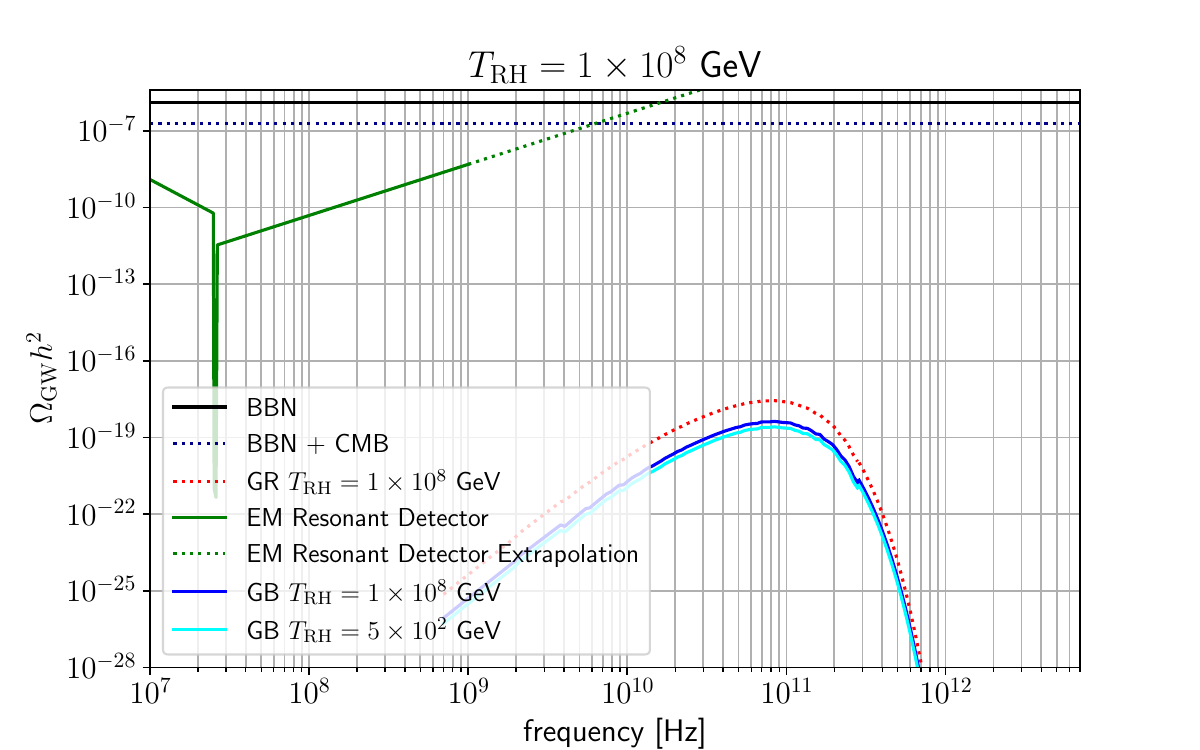}   
    \includegraphics[width=7.3cm]{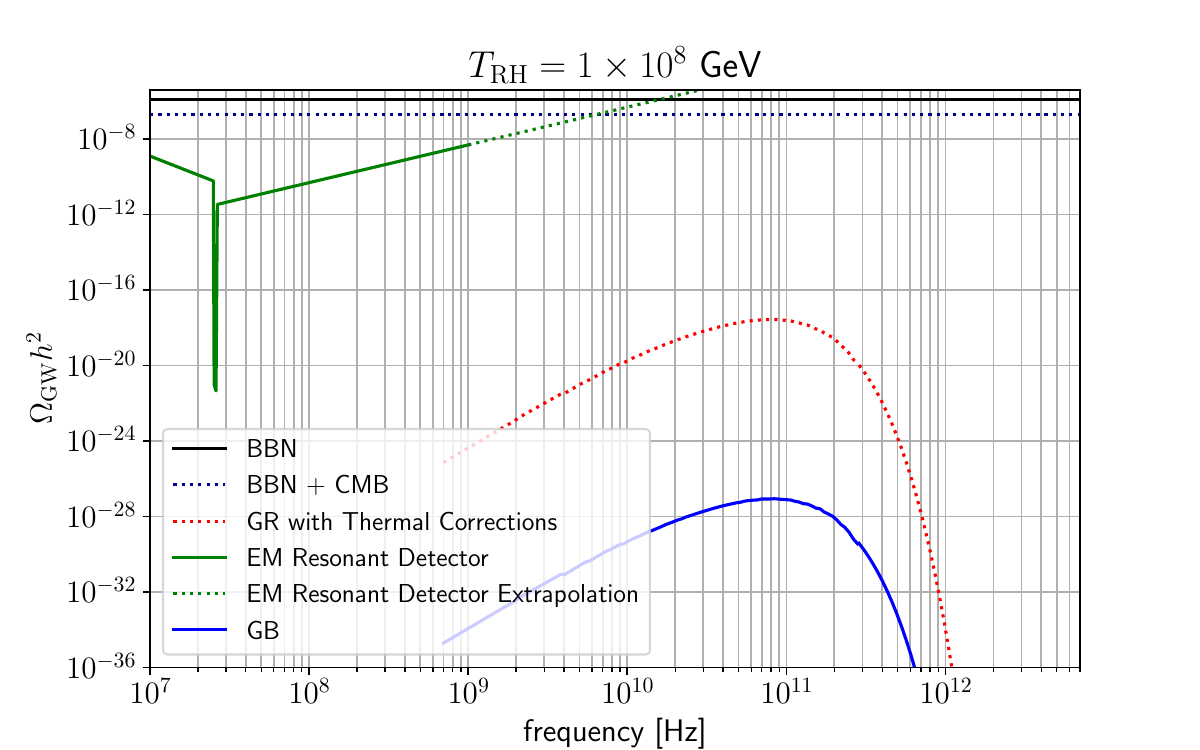}
    \caption{Same as Fig.~\ref{fig:GWconstEx1}, but for all of these cases 
    $\tilde{\alpha} = 1 \, {\rm km^2}$. For the first panel $\gamma=-1$, second panel 
    $\gamma=1$, third panel $\gamma=-5$ and fourth panel $\gamma=5$  .}
    \label{fig:GWconstEx2}
\end{figure}

This is confirmed by the plots of Fig.~\ref{fig:GW_signal_alpha_gamma}, where the color code shows the peak value of the expected GW stochastic background for $T_{\rm max}=10^{16}$ GeV normalised to the BBN upper limit. Specific examples of the expected GW stochastic frequency spectra for 
$\tilde{\alpha}=\pm 1$km$^2$, $\gamma = \pm 1$ and $\gamma = \pm 5$ (corresponding to each of the flows numbered from $\circled{1}$ to $\circled{8}$ in Fig.~\ref{fig:critical_points})  are also provided for reference in Fig.~\ref{fig:GWconstEx1}  ($\tilde{\alpha}>0$) and~\ref{fig:GWconstEx2} ($\tilde{\alpha}<0$) in the case when $T_{\rm max}=10^8$ GeV. The same figures also show the BBN bound and the prospective constraints (extrapolated to higher frequencies) from EM cavities of Ref.~\cite{Herman:2022fau}. Indeed, the normalized GW peak value in Fig.~\ref{fig:GW_signal_alpha_gamma} largely exceeds unity whenever the asymptotic equation of state corresponds to slow roll ($w=-1/3$), indicating that in such regions of the parameter space it is possible to set an upper bound on $T_{\rm max}\ll10^{16}$ GeV. This is confirmed by Fig.~\ref{fig:Tmax_GW_alpha_gamma}, where the cyan shaded areas for $\tilde{\alpha}<0$ indicate the $T_{\rm max}$ bound obtained from the BBN constraint on the GW background, while the other regions (where the GW signals are too weak to obtain any bound) 
show the maximum possible temperature for the production of the GW signal (which corresponds to the highest temperature for which $\rho_{\rm rad}$ dominates the energy density of the Universe). 

\begin{figure*}[htb!]
\centering
\includegraphics[width=7.3cm]{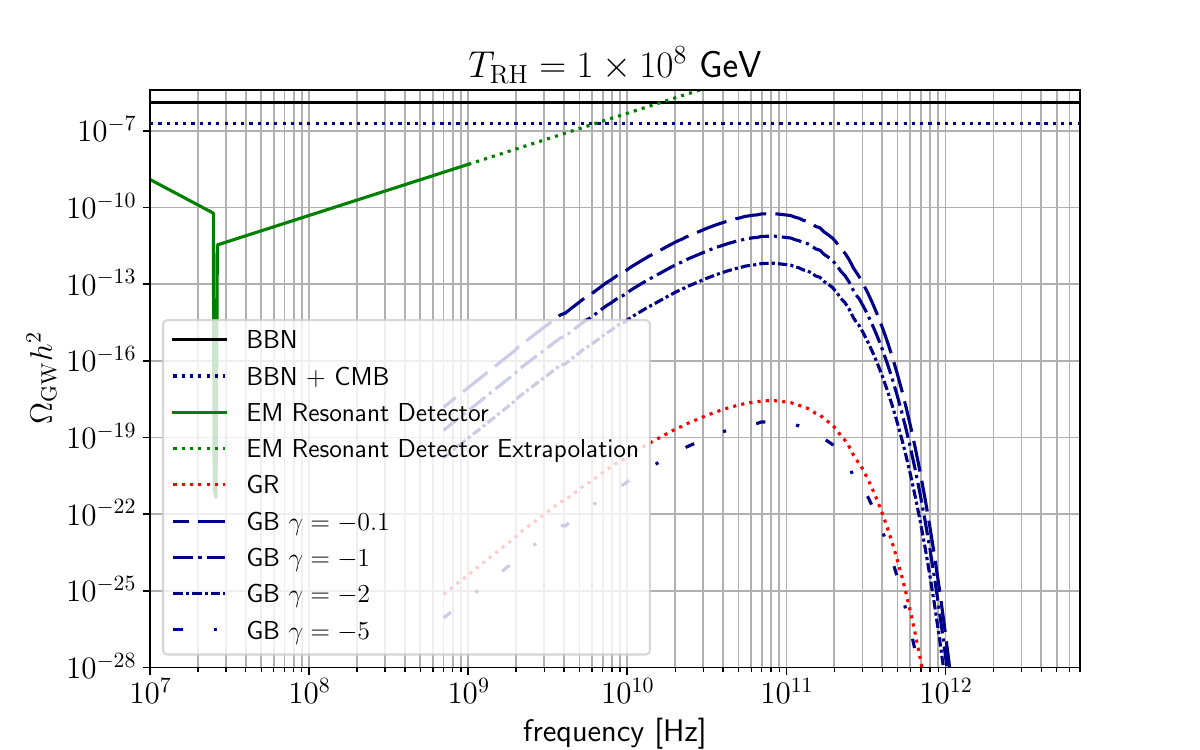}
\caption{Same as in Figs.~(\ref{fig:GWconstEx1}-\ref{fig:GWconstEx2})  for 
$\tilde{\alpha} = 1 \, {\rm km^2}$ and different values of $\gamma<0$. The signal is enhanced as $|\gamma|\rightarrow 0$ because the metastable slow--roll solution with $w$=-1/3 lasts longer (see Fig.~\ref{fig:eos_neg_gammas}).}
\label{fig:GW_alpha_positive_gamma_negative}
\end{figure*}

\begin{figure*}[htb!]
\centering
\includegraphics[width=7.49cm,height=6.3cm]{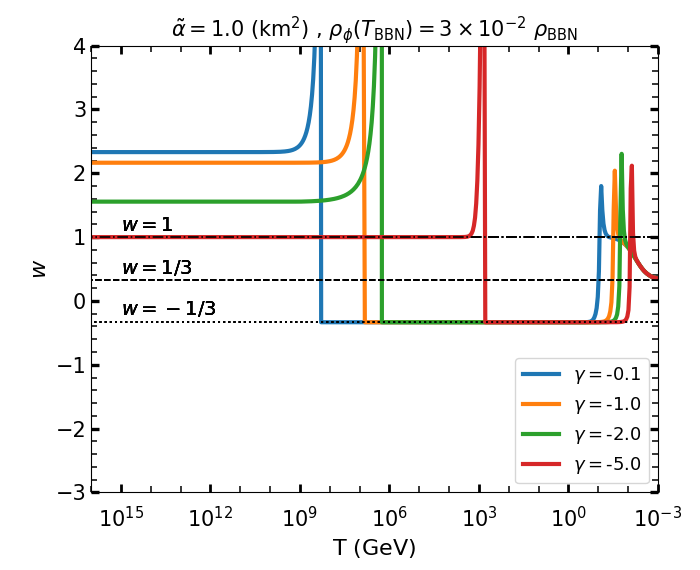}
\caption{Evolution of the equation of state for $\tilde{\alpha} = +1 \, {\rm km^2}$, $\rho_\phi(T_{\rm BBN})$ = $3\times10^{-2}\rho_{\rm BBN}$, for different negative values of $\gamma$. As $|\gamma|\rightarrow$0 the system follows the metastable solution $w=-1/3$ for a larger interval of temperatures before jumping to a different regime, implying an increasing GW stochastic background (see Fig.~\ref{fig:GW_alpha_positive_gamma_negative}).} 
\label{fig:eos_neg_gammas}
\end{figure*}

In particular, in agreement with the discussion of Section~\ref{sec:flows}, a bound on $T_{\rm max}$ can be obtained for $\dot{\phi}(T_{\rm BBN})=0$ in the full parameter space with $\tilde{\alpha}<0$. However, for $\dot{\phi}(T_{\rm BBN})>0$ the correspondence between the slow-roll regions in Fig.~\ref{fig:eos_alpha_gamma} and the high GW signal domains in Figs.~\ref{fig:GW_signal_alpha_gamma} and~\ref{fig:Tmax_GW_alpha_gamma} is no longer exact. Specifically, in Figs.~\ref{fig:GW_signal_alpha_gamma} and~\ref{fig:Tmax_GW_alpha_gamma} some  regions with a high GW signal (or at least higher than in Standard Cosmology) are observed for configurations where $w\ne-1/3$. Moreover, at variance with the equation--of--state domains of Fig.~\ref{fig:eos_alpha_gamma}, the regions of Fig.~\ref{fig:GW_signal_alpha_gamma} show some dependence on the $\tilde{\alpha}$ parameter. The reason of such mismatch is simply because the GW production process depends on the whole expansion history of the Universe, and not only on its asymptotic behaviour at very large temperatures. In particular, as discussed in Section~\ref{sec:flows}, for  $\dot{\phi}(T_{\rm BBN})>0$   there exist specific situations for which the boundary conditions at BBN imply $z_{\rm BBN}>0$ while the only available attractor is at $z<0$ and is not slow-roll (instead, it is either kination at $(1,z\rightarrow 0^-)$ or $(x_{\rm fp},z_{\rm fp})$).  In such situations the boundary conditions at BBN ``go against the flow”, and in order to cross the origin the system needs to follows the $z$ axis with $w=-1/3$ for a while before eventually jumping to the real attractor at higher temperatures (see paths $\circled{6}$ and $\circled{8}$ in Fig.~\ref{fig:critical_points}). In these cases a metastable regime is achieved where, although the asymptotic behaviour at high temperature is not slow-roll, the system does follow the slow-roll equation of state in some interval of temperatures, implying an enhancement of the GW signal compared to the case of standard Cosmology, albeit not at the same level of the cases when slow-roll is the stable attractor at indefinitely high $T$. This is the explanation of the region of intermediate values of the GW signal that is observed in Fig.~\ref{fig:GW_signal_alpha_gamma} for $\dot{\phi}(T_{\rm BBN})>0$, $\tilde{\alpha}>0$ and $\gamma<0$. The fact that such domain extends up to some maximal value of $|\gamma|$ can be explained in a qualitative way by noting that for $\gamma\rightarrow 0$ the attractor $(x_{\rm fp},z_{\rm fp})$ moves away to $(+\infty,-\infty)$ along the $x+z=1$ line, implying that for smaller values of $|\gamma|$ the corresponding flow (see path $\circled{8}$ in Fig.~\ref{fig:critical_points}) detaches ``later" from the $z$ negative axis, and the metastable slow-roll solution responsible for the signal enhancement lasts longer. As a consequence for $\tilde{\alpha}<0$ and $\gamma<0$ the GW signal is a decreasing function of $|\gamma|$, as shown explicitely in Fig.~\ref{fig:GW_alpha_positive_gamma_negative}.

\begin{figure*}[htb!]
\centering
\includegraphics[width=9.5cm,height=7.5cm]{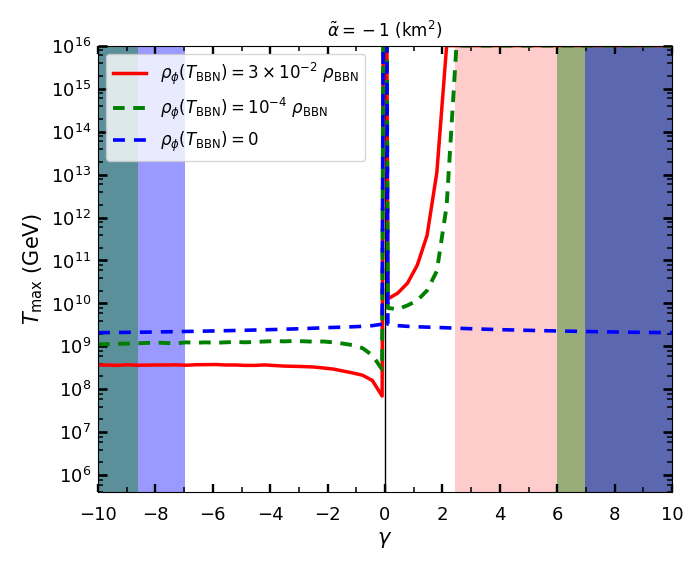}
\caption{GW upper bounds on the maximum temperature of the Universe as a function of $\gamma$ and for $\tilde{\alpha} = -1 \, {\rm km^2}$. In the caption $\kappa$ =1. The curves represent the bounds obtained by imposing that the peak value of the GW stochastic background does not exceed the BBN upper limit ($\Omega_{\rm GW, lim}h^2 \simeq 10^{-6}$), while the shaded areas are excluded by the observation of GW from binary mergers.
The red, green and blue curve and shaded areas correspond to $\rho_\phi(T_{\rm BBN})$ = $3\times10^{-2}\rho_{\rm BBN}$, $10^{-4}\rho_{\rm BBN}$ and $0$, respectively.}
\label{fig:TRH_uplim_vs_gamma}
\end{figure*}

\begin{figure*}[htb!]
\centering
\includegraphics[width=7.49cm,height=6.2cm]{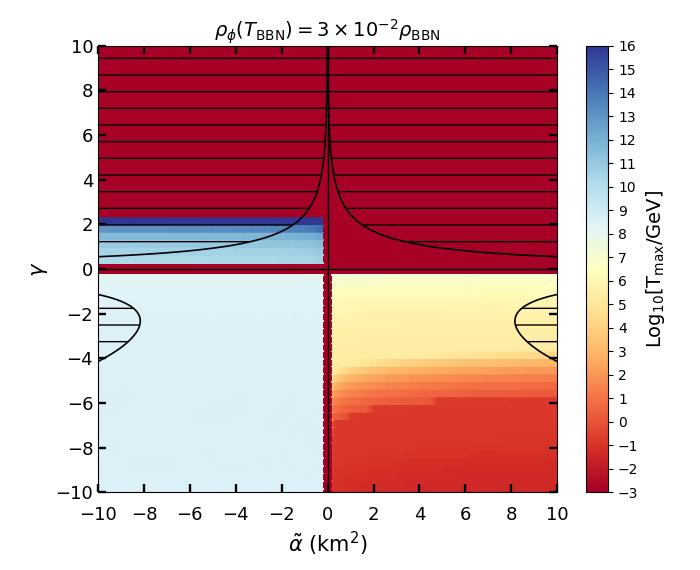}
\includegraphics[width=7.49cm,height=6.2cm]{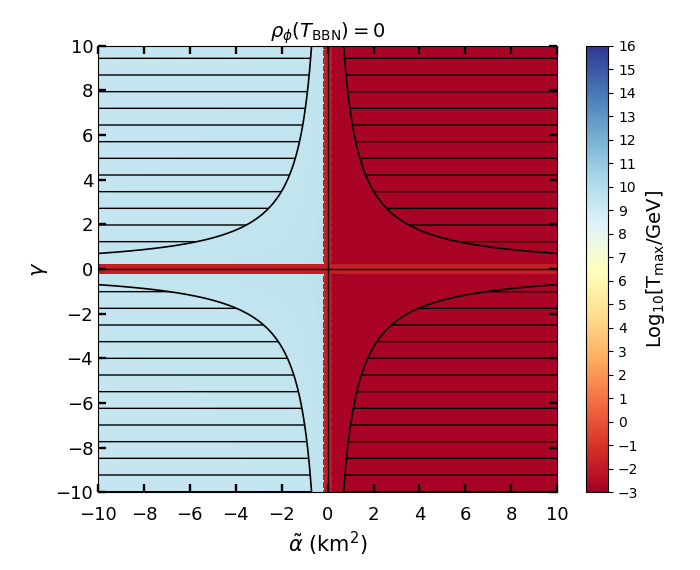}
\includegraphics[width=7.49cm,height=6.2cm]{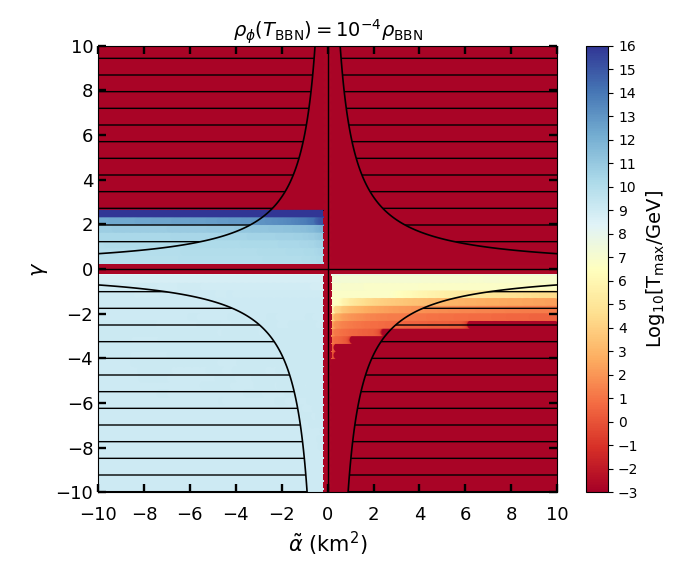}
\caption{The color codes show the variations of $T_{\rm max}$ in the $\tilde{\alpha} - \gamma$ plane 
for different values of $\rho_\phi(T_{\rm BBN})$: $3\times10^{-2}\rho_{\rm BBN}$ (top left), $0$ (top right) and $10^{-4}\rho_{\rm BBN}$ (bottom). In each plot, the cyan region for $\tilde{\alpha} < 0$ 
indicates the upper bound on the maximum temperature of the Universe obtained from the BBN constraint on GW background, while the other regions (where the GW signals are too weak to obtain any bound) 
show the maximum possible temperature for the production of the GW signal. 
The hatched areas are ruled out by the detection of GW from compact binary mergers.}
\label{fig:Tmax_GW_alpha_gamma}
\end{figure*}

To summarize the discussion of this Section, the stochastic GW background can set a meaningful bound on $T_{\rm max}<10^{16}$ GeV only for $\tilde{\alpha}<0$, when the slow-roll attractor solution is achieved ($w=-1/3$). In particular when $\dot{\phi}(T_{\rm BBN})=0$ 
the flow of the autonomous differential equations always moves from $(0,0)$ directly toward $z<0$ and the system quickly achieves the slow--roll asymptotic behaviour, so that such bound is possible for any value of $\gamma$.  On the other hand when $\dot{\phi}(T_{\rm BBN})>0$ the system originates from $(x_{\rm BBN},z_{\rm BBN})$ with $z_{\rm BBN}\ne 0$. In this case, while for $\gamma<0$ one has $z_{\rm BBN}<0$, and a situation similar to the $\dot{\phi}(T_{\rm BBN})=0$ case is obtained, when $\gamma>0$ one has $z_{\rm BBN}>0$. In this case when $\gamma<\sqrt{6\kappa}$ slow roll remains the only attractor, so that the system eventually reaches it after crossing the origin. However, when $\gamma>\sqrt{6\kappa}$ a second attractor appears at $(1, z\rightarrow 0^+)$ which deviates the flow away from slow--roll toward kination. As a consequence, for $\gamma\gtrsim\sqrt{6\kappa}$ the expected GW stochastic background is suppressed and no bound on $T_{\rm max}$ can be obtained. 
Explicit examples of the bounds on $T_{\rm max}$ that can be obtained for $\tilde{\alpha}<0$ are represented by the curves in Fig.~\ref{fig:TRH_uplim_vs_gamma}, where $\tilde{\alpha}=-1\,\mbox{km}^2$. In such Figure the dashed blue line shows that a bound of order $T_{\rm max}\lesssim 10^9$ GeV can be obtained at all $\gamma$ values when 
$\rho_\phi(T_{\rm BBN})=0$, while for $\rho_\phi(T_{\rm BBN})=3\times 10^{-2}\rho_{\rm BBN}$ (red solid line) and $\rho_\phi(T_{\rm BBN})=10^{-4}\rho_{\rm BBN}$ (green dashed line) the bounds quickly disappear for $\gamma\gtrsim\sqrt{6\kappa}$.

In figures \ref{fig:GW_signal_alpha_gamma} and \ref{fig:Tmax_GW_alpha_gamma} the shaded
regions are combined with hatched areas ruled out by the merger events discussed in Section~\ref{sec:GW_binaries} and obtained using Eq.~(\ref{eq:GW_constraints_2}). 
From such figures one can notice that the GW bound from binaries allows to constrain the dEGB parameter space in some of the regions ($\tilde{\alpha}>0$ and $\tilde{\alpha}<0$, $\gamma>\sqrt{6\kappa}$) where the high--temperature asymptotic equation of state does not correspond to slow roll ($w=-1/3$) and the GW stochastic background is suppressed. Moreover, such bound does not depend on $T_{\rm RH}$. The complementarity between the two bounds is also observed for $\tilde{\alpha}<0$ and $\gamma>0$, where the bound from the GW stochastic background relaxes quickly as $\gamma\rightarrow\sqrt{6\kappa}$ (see  Fig.~\ref{fig:TRH_uplim_vs_gamma}).

\section{Conclusions}
\label{sec:conclusions}

In spite of a high degree of non linearity and phenomenological complexity at low temperature, Cosmology in the dilaton-Einstein-Gauss-Bonnet (dEGB) scenario of modified gravity exhibits only very few asymptotic behaviours at large temperature, characterized by only a few values of the equation of state $w=p/\rho$ (with $\rho$ and $p$ the energy density and pressure of the Universe, respectively), when the model is required to reproduce Standard Cosmology at BBN. In the present paper we have provided a transparent and systematic discussion of such high temperature asymptotic behaviours, that were first observed in the numerical analysis of Ref.~\cite{GB_WIMPS_sogang}.

By re--expressing the Friedmann equations of the model (Eqs.(\ref{Eqfrw1}, \ref{Eqfrw2}, \ref{Eqfrw3})) in terms of a set of autonomous differential equations (Eqs.~(\ref{eq:x_prime}, \ref{eq:z_prime})) we found that the such asymptotic behaviours have a clear interpretation in terms of a pattern of attractors (see Table~\ref{table:critical_points} and Fig.~\ref{fig:critical_points}). Specifically, in terms of the dEGB parameters $\tilde{\alpha}$ and $\gamma$ and of the value of the kinetic energy of the scalar field at BBN $\rho_{\phi}(T_{\rm BBN})$,  we found that:

\begin{itemize}
    \item The model presents only three attractors at high temperature : slow-roll ($w=-1/3$), kination ($w=1$) and fast-roll ($1<w<7/3)$. 
    \item The way such attractor solutions are approached depends on only three conditions: (i) $\tilde{\alpha}>0$ or  $\tilde{\alpha}<0$; (ii) $\gamma>0$ or  $\gamma<0$; (iii) $|\gamma|<\sqrt{6\kappa}$ or $|\gamma|>\sqrt{6\kappa}$. The combination of these three dichotomies generates eight possible paths for the evolution of the system at high temperature, represented in Fig.~\ref{fig:critical_points} and in Table~\ref{table:critical_points} with the numbers from $\circled{1}$ to $\circled{8}$. When $\rho_{\phi}(T_{\rm BBN})=0$ only half of the paths are followed: $\circled{2}$, $\circled{4}$, $\circled{5}$ and $\circled{7}$. For our conventional choice $\dot{\phi}(T_{\rm BBN})>0$ such paths are also followed when $\rho_{\phi}(T_{\rm BBN})\ne0$ if $\tilde{\alpha}\gamma<0$, leading to phenomenological results 
    very similar to the corresponding ones for $\dot{\phi}(T_{\rm BBN})=0$. 

    \item The correspondence between path and asymptotic $w$ is the following: $w=-1/3$ (slow--roll) for $\circled{1}$, $\circled{2}$ and $\circled{4}$; $w=1$ (kination) for $\circled{3}$, $\circled{7}$ and $\circled{8}$;  $1<w<7/3$ (fast--roll) for $\circled{5}$ and $\circled{6}$.

    \item The effect of the GB term on the evolution of the Friedmann equations can be parameterized in term of the additional component $\rho_{\rm GB}$ to the energy density of the Universe (see Eq.~(\ref{Eqfrw1})). Such component is not physical, since it can be negative. Paths $\circled{1}$, $\circled{6}$ and $\circled{8}$ are characterized by the fact that the boundary conditions imply $\rho_{\rm GB}(T_{\rm BBN})>0$, while the only available attractor at high temperature has $\rho_{\rm GB}<0$. This requires a change in the sign of $\rho_{\rm GB}$ and, so of $\dot{\phi}$, that makes the approaching path to the attractor longer (the system need to ``go against the flow", see discussion in Section~\ref{sec:results}).  Such cases, which are only possible for $\dot{\phi}(T_{\rm BBN})\ne 0$, are characterized by the emergence of meta-stable solutions, where the system follows an equation of state for a while at intermediate temperatures  before eventually jumping to the real attractor at higher temperatures. Specifically, for $\circled{1}$ the system follows kination ($w=1$) before jumping to slow--roll ($w=-1/3$), while for $\circled{6}$ and $\circled{8}$ the system follows slow-roll ($w=-1/3$) before jumping to fast roll ($1<w<7/3$). We notice that in the analysis of Ref.~\cite{GB_WIMPS_sogang} the evolution of the  dEGB Friedmann equations was limited to the temperatures relevant for WIMP thermal decoupling, and the case $\tilde{\alpha}$ = 1 km$^2$, $\gamma=1$, $\rho_{\phi}(T_{\rm BBN})=3\times 10^{-2}\rho_{\rm BBN}$ (corresponding to the path $\circled{6}$) was mistakenly assumed to have the slow-roll asymptotic behaviour, missing the transition to fast roll at higher temperatures.

    \item Compared to standard Cosmology, the regions of the dEGB parameter space corresponding to a slow-roll asymptotic equation of state ($w=-1/3$) imply a strong enhancement of the expected Gravitational Wave stochastic background produced by the primordial plasma of relativistic particles. This already allows to use the bound from BBN to put sensible bounds on $T_{\rm RH}\simeq 10^8-10^9$ GeV $\ll 10^{16}$ GeV. Interestingly, such enhancement is possible because slow--roll in dEGB allows to have and epoch when the energy density $\rho_{\rm rad}$ of the relativistic plasma dominates the energy of the Universe while at the same time the rate of dilution with $T$ of the total energy density is slower than what usually expected during radiation dominance (specifically, the energy density of the Universe is dominated by $\rho_{\rm rad}$, $|\rho_{\rm GB}|\propto T^4$ while at the same time $\rho_{\rm rad}+\rho_{\rm GB}\propto T^2$, with a large cancellation between $\rho_{\rm rad}$ and $\rho_{\rm GB}$).

    \item The GW bound from binaries allows to constrain the dEGB parameter space in some of the regions where the high--temperature asymptotic equation of state does not correspond to slow roll ($w=-1/3$) and the GW stochastic background is suppressed. Moreover, such bound does not depend on $T_{\rm RH}$.
\end{itemize}

We conclude by pointing out that we analyzed dEGB cosmology by fixing the boundary conditions of the Friedmann equations at BBN and run them to higher temperatures, i.e. backward in time. On the other hand, the physical (causal) evolution of the Universe is obtained by reversing the flow, starting from some initial conditions at high temperatures. 
In such case the origin of the $x-z$ plane, which corresponds to standard radiation dominated $\Lambda$CDM Cosmology, is technically not a critical point of the set of autonomous equations (\ref{eq:x_prime}, \ref{eq:z_prime}). Indeed, the flow can cross the origin (see for instance the paths $\circled{1}$, $\circled{3}$, $\circled{6}$ and $\circled{8}$ of Fig.~\ref{fig:critical_points}) so that $x^{\prime},z^{\prime}\ne 0$ for $(x,z)$ = $(0,0)$. Nevertheless, when the Friedmann equations are run below BBN one observes that both $\rho_{\rm GB}$ and $V_{\rm GB}^{\prime}$ are suppressed by their dependence on $H$, while the friction term in the scalar field equation~(\ref{Eqfrw3}) eventually drives $\dot{\phi}$ to zero. This implies that, starting from some high temperature $T_{\rm max}$, when the system is evolved for a large--enough number $\Delta N$ of e--foldings it eventually gets arbitrarily close to the origin at lower temperatures (without formally reaching it) and Standard Cosmology is recovered. From this perspective the BBN bound rules out high--temperature configurations for which $\Delta N$ is too large, and Standard Cosmology is achieved too late (specifically, below $T_{\rm BBN}$). By fixing the boundary conditions at BBN the high--temperature solutions found in the present paper reach standard Cosmology fast enough by construction, but one may wonder if they require some level of fine tuning i.e. if $\Delta N$ is sensitive to them.
A discussion of this aspect would require a dedicated and systematic numerical exploration of the Friedmann equations that goes beyond the scope of the present paper. It would be an important aspect to consider when discussing which inflationary and reheating models within dEGB can lead to Cosmologies compatible to observation.

\section*{Acknowledgements}
This research was supported by the National Research Foundation of Korea (NRF) funded by the Ministry of Education through the Center for Quantum Space Time (CQUeST) with grant number
2020R1A6A1A03047877 and by the Ministry of Science and ICT with grant
numbers RS-2023-00241757. The  work of HC. Lee and L. Velasco-Sevilla is partially supported by the National Research Foundation of Korea (NRF) grant, RS-2023-00273508, B.-H. Lee (2020R1F1A1075472) and W. Lee (2022R1I1A1A01067336). BHL thanks APCTP and KIAS, where part of this work was done, for the warm hospitality. L. Yin is supported by an appointment to the YST Program at the APCTP through the Science and Technology Promotion Fund and Lottery Fund of the Korean Government.

\appendix

\section{Thermal corrections} 
\label{app:thermalcorr}
In order to identify with ease the thermal corrections computed in \cite{Ghiglieri:2020mhm}, we collect the expressions of \cite{Ghiglieri:2015nfa} and \cite{Ghiglieri:2020mhm} rewritten in \cite{Ringwald:2020ist} with different notation. 
The function $\eta(T,\hat{k})$ can be identified with $R(T,\hat{k})$, from Eq. (6.5)  of \cite{Ghiglieri:2015nfa} and $\Gamma$ from \cite{Ghiglieri:2020mhm} as follows:
\bea
R(\hat{k},T)= 2 k \Gamma n_B(\hat{k})= 32 \pi \frac{T^4}{\Mp^2}  \eta(\hat{k},T),
\eea
where the function $\Gamma$ is the main subject of computation in \cite{Ghiglieri:2020mhm}. Hence $\eta(T,\hat{k})$ is given by
\bea
\label{eq:etaGamma}
\eta(\hat{k},T) =\frac{1}{16\pi} \frac{k}{T^4} \Mp^2 \Gamma n_B(\hat{k}).
\eea
We write the contributions to $\Gamma$ with the same notation used in Eq.~(2.8) of \cite{Ghiglieri:2020mhm} for the Euclidean correlator $G^E_{12;12}$ as
\bea
\Gamma(k)=\frac{16\pi \mathrm{Im}\left(G^E_{12;12}|_{k_n\rightarrow -i[k+i0^+]}\right)}{k \Mp^2},
\eea
where $G^E_{12;12}$ is a function of the 1-loop diagrams and 2-loop diagrams (for which for example $\Phi_{a(b)}$  represents the 2-loop diagram describing the coupling of the particle of type a to $T_{\mu\nu}$ where a particle of type $b$ appears in the loop). Hence we write
\bea
\label{eq:EuclideanG}
\mathrm{Im}\left(G^E_{12;12}|_{k_n\rightarrow -i[k+i0^+]}\right)=\sum  C_{\Phi_x} \hat{\Phi}_x, 
\eea
where $C_{\Phi_x}$ is a numerical coefficient and
$\hat{\Phi}_x$ is a loop function. Both can be read off from Eq.~(2.6) of \cite{Ghiglieri:2020mhm} after taking the limit of \eq{eq:EuclideanG}.
In this way, we can identify all the components of 
$\eta(\hat{k},T)$ of \eq{eq:etaHTL} and \eq{eq:etaThermal} since then, \eq{eq:etaGamma} becomes
\bea
\eta(\hat{k},T)= \frac{n_B(\hat{k})}{T^4} \sum  C_{\Phi_x} \hat{\Phi}_x.
\eea
Taking $D=4$, $N_g=3$ in Eq.~(2.8) of \cite{Ghiglieri:2020mhm}, we have
\bea
\eta_{\rm{HTL}}(\hat{k},T)=\frac{n_B(\hat{k})}{T^4}  \frac{1}{4}  \sum_{i=1,2,3} (2+ N^i_c C_F^i) \hat{\Phi}^i_g,
\eea
where $i=1,2,3$ runs over the SM gauge factors $C_F=(N^2_c-1)/2/N_c$, $N_c$ being of course different for the different SM gauge groups, and 
\bea
\label{eq:ThersemE}
\eta^T(\hat{k},T)&=& \frac{1}{4}\frac{n_B(\hat{k})}{T^4} 
\left\{
\left[(3 g_2^2 + 12 g_3^2) \right] \hat{\Phi}_{g(g)}+(g_1^2+3g_2^2) \left[\hat{\Phi}_{s(g)}+ \hat{\Phi}_{g(s)} + \hat{\Phi}_{s|g}\right]\right.\nonumber\\
& &
+ (5 g_1^2+ 9g_2^2 +24 g_3^2)\left[\hat{\Phi}_{f(g)}+ \hat{\Phi}_{g(f)} + \hat{\Phi}_{f|g}\right]\nonumber\\
&  &
\left.
+ (3 |y_t|^2+ 3|y_b|^2+|y_\tau|^2)\left[\hat{\Phi}_{s(f)}+ \hat{\Phi}_{f(s)} + \hat{\Phi}_{s|f}\right]
\right\}.
\eea
Therefore, we can identify the functions $\eta_{gg}(\hat{k})$, $\eta_{sg}(\hat{k})$, $\eta_{fg}(\hat{k})$ and $\eta_{sf}(\hat{k})$, for example 
\bea
\eta_{gg}(\hat{k})=\frac{1}{4} \frac{n_B(\hat{k})}{T^4} \left[\hat{\Phi}_{s(g)}+ \hat{\Phi}_{g(s)} + \hat{\Phi}_{s|g}\right].
\eea
The important point is that all of the functions $\eta_{gg}(\hat{k})$ and of course equivalently $\hat{\Phi}$ have contributions from t and s channels.  In this way, we can write
\bea
\left[\hat{\Phi}_{s(g)}+ \hat{\Phi}_{g(s)} + \hat{\Phi}_{s|g}\right]= H^t_{+} + H^s_+, 
\eea
where the functions $H^t_{+}$ and $H^s_{+}$ can be read off from Eqs.~(2.60), (2.61), (2.71) with the coefficients of Eqs.~(2.72-2.75) of \cite{Ghiglieri:2020mhm}. Hence, we get for all  of the functions appearing in \eq{eq:ThersemE} the following expressions:
\bea
\eta_{gg}(\hat{k})&=& \frac{1}{4} \frac{n_B(\hat{k})}{T^4}\left(H^t_{+}(\hat{k}) + H^s_{+}(\hat{k}) \right),\nonumber\\
\eta_{sg}(\hat{k})&=&\frac{1}{4} \frac{n_B(\hat{k})}{T^4}\left[\frac{1}{4} \left( H^t_{+}(\hat{k}) + H^s_{+}(\hat{k}) \right] \right],\nonumber\\
\eta_{sf}(\hat{k})&=& \frac{1}{4} \frac{n_B(\hat{k})}{T^4}
\left[ K^t(\hat{k})+K^s(\hat{k}) \right],\nonumber\\
\eta_{fg}(\hat{k})&=& \frac{1}{4} \frac{n_B(\hat{k})}{T^4}
\left[-H^t_{-}-H^s_{-} \right],
\eea
where in the notation of \cite{Ringwald:2020ist}, this translates into 
\bea
H^t_{\pm}(\hat{k})=\frac{\mathcal{I}_{\pm}(\hat{k}) \, T^4}{-16 n_B(\hat{k})}, &&
H^s_{\pm}(\hat{k})=\frac{\mathcal{J}_{\pm} (\hat{k})\, T^4}{-16 n_B(\hat{k})},\nonumber\\
K^t(\hat{k})= \frac{\mathcal{K}(\hat{k}) T^4}{16 n_B(\hat{k})}, &&
K^s(\hat{k})= \frac{\mathcal{L}(\hat{k}) T^4}{16 n_B(\hat{k})},
\eea
and hence 
\bea
H^t_{\pm}(\hat{k})&=&\frac{4 T^4}{(4\pi)^3 \hat{k}} 
\int_{-\infty }^{\hat{k}} dx \, \int_{|x|}^{2\hat{k}-x} dy 
\left[( 1+ n_B(x)+n_B(\hat{k}-x))(y^2-x^2)\right.\nonumber \\
& & -\frac{2}{3} L_1^{\pm} + \frac{(y^2-3(x-2\hat{k})^2)(12 L_3^{\pm} + 6 y L_2^{\pm} + y^2L_1^{\pm} )}{6 y^4} \nonumber\\
& & \left. + \frac{(1\pm 3)x {\hat{k}}^2 \pi^2}{y^4} (y^2-x^2)   \right]. \label{eq:phi1gg}
\eea
Here, we note the missing ``$x$'' factor in Eq.~(A.2) of  \cite{Ringwald:2020ist}, which we include here in the last line of \eq{eq:phi1gg}. This factor simply follows from Eq.~(2.61) of \cite{Ghiglieri:2020mhm}, coming from the divergent part of the t-channel contribution. The functions $L_i^{\pm}$ are given in Eqs. (A.4) of \cite{Ringwald:2020ist} and can be identified with Eqs.~(2.57-2.59) of \cite{Ghiglieri:2020mhm} with the appropriate coefficients given in Eqs. (2.72-2.75) of \cite{Ghiglieri:2020mhm}.
For the s-channel contribution we have
\bea
H^s_{\pm}(\hat{k})&=& \frac{4 T^4}{(4\pi)^3 \hat{k}} \int_{\hat{k}}^{\infty}dx\int_{|2\hat{k}-x|}^{x} dy\,(n_B(x-\hat{k})-n_B(x))(y^2-x^2)\!\left(\frac{1}{3}(2M^{\pm}_1(x,y){-}y)\right.\nonumber\\
& -&\left.\frac{(y^2-3(x-2\hat{k})^2)}{6 y^4}(12M^{\pm}_3(x,y)+6yM^{\pm}_2(x,y)+y^2M^{\pm}_1(x,y)) \right.\nonumber\\
&
+ &\left. (x-2\hat{k})\left[ \hat{M}_2^{\pm}(x,y)-2 y \hat{M}_1^{\pm}(x,y)   \right]  \right), \label{eq:phigg2}
\eea
where $x=q_0/T$, $y=q/T$, $q_0$ and $q$ the quantities (momenta) appearing in Eq.~(2.71)\cite{Ghiglieri:2020mhm}  $\hat{M}_2^+=\frac{2}{T^2}(\hat{L}^-_2-\hat{L}^+_2)=0$ and $\hat{M}_1^+=\frac{2}{T}(\hat{L}^-_1-\hat{L}^+_1)=0$. The functions $M_i^{\pm}$ are given in Eq.~(A.4) of \cite{Ringwald:2020ist} and can be extracted from Eqs. (2.68-2.70) with the appropriate coefficients of Eqs. (2.72-2.75) of \cite{Ghiglieri:2020mhm}. 
For the functions $\mathcal{K}$ and $\mathcal{L}$ we find expressions in agreement with those in Eq. (A.2) of \cite{Ringwald:2020ist}.

\newpage


\end{document}